\documentstyle[12pt,rotating]{article}

\input{psfig}

\title{\begin{flushright}
{\normalsize NUC-MINN-97/1-T\\
March 1997 \\}
\end{flushright}
\vspace*{0.3in}
{\bf LINEAR EXTRAPOLATION OF ULTRARELATIVISTIC NUCLEON-NUCLEON
SCATTERING TO NUCLEUS-NUCLEUS COLLISIONS}}
\author{{\bf Sangyong Jeon}$^{\dag}$ and {\bf Joseph Kapusta}$^{\ddag}$\\
  {\it School of Physics and Astronomy}\\
  {\it University of Minnesota}\\ {\it Minneapolis, MN 55455}}

\date{}

\begin{document}

\maketitle

\begin{center}
Abstract
\end{center}

\noindent
We use a Glauber-like approach to describe very energetic
nucleus-nucleus collisions as a sequence of binary nucleon-nucleon
collisions.  No free parameters are needed: all the information comes
from simple parametrizations of nucleon-nucleon collision data.
Produced mesons are assumed not to interact with each other or with
the original baryons.  Comparisons are made to published experimental
measurements of baryon rapidity and transverse momentum distributions,
negative hadron rapidity and transverse momentum distributions,
average multiplicities of pions, kaons, hyperons, and antihyperons, and
zero degree energy distributions for sulfur-sulfur collisions at
200 GeV$/c$ per nucleon and for lead-lead collisions at 158 GeV$/c$
per nucleon.  Good agreement is found except that
the number of strange particles produced, especially antihyperons, is
too small compared with experiment.  We call this model LEXUS:
it is a baseline linear extrapolation of ultrarelativistic
nucleon-nucleon scattering to heavy ion collisions.

\vspace{.5in}

\noindent
$^{\dag}$ Jeon@nucth1.spa.umn.edu\\
\noindent
${}^{\ddag}$ Kapusta@physics.spa.umn.edu

\newpage

\section{Introduction}

The program to study the properties of quark-gluon plasma at high
energy density is in high gear \cite{qm} with the construction of the
Relativistic Heavy Ion Collider (RHIC) at Brookhaven National Laboratory
(BNL) well underway and due to be completed in 1999.  In this machine
counter-rotating beams of gold nuclei with an energy of 100 GeV per
nucleon will be collided.  The Large Hadron Collider (LHC) at CERN
will be completed around the year 2005; it will allow for the collision
of counter-rotating beams of lead nuclei at about 1.5 TeV per nucleon,
but will not be dedicated to heavy ion physics as RHIC will be.
Since 1986 experiments have been performed at CERN's SPS accelerator
with beams of oxygen and sulfur at 200 GeV$/c$ per nucleon and,
lately, lead at 158 GeV$/c$ per nucleon, striking fixed targets.
In the same
time interval similar beams have been available at BNL's AGS accelerator
at the lower energies of 10 to 14.6 GeV per nucleon.
It is almost universally accepted that the proper treatment of
collisions at RHIC and LHC must involve the quark and gluon degrees
of freedom.  At the AGS hadronic degrees of freedom probably suffice
(but see \cite{nucleate}).  The jury is still out concerning collisions at
the SPS.

It is oftentimes heard at conferences and workshops that there is a
need for a baseline calculation of what one would expect at the
above heavy ion accelerators if there was no new physics; that is,
a {\it linear} extrapolation of nucleon-nucleon collisions to
nucleus-nucleus collisions.  Construction of such a working
model is the goal of this paper.  Actually it is not so obvious how
to make such a linear extrapolation.  Nucleon collisions produce mesons,
and these mesons can collide with other nucleons and mesons, producing
an interesting cascade of hadrons.  We do not consider such a cascade
as being a linear extrapolation.  The collective excitations of
such a system are not necessarily trivial, nor is it a simple matter
to compute or measure all the hadronic cross sections needed to keep
track of this cascade.  Many such models already exist: ARC \cite{arc},
RQMD \cite{rqmd}, VENUS \cite{venus}, FRITIOF \cite{fritiof},
PYTHIA \cite{pythia}, and QGSM \cite{qgsm}
being among the most frequently applied to experimental data.

Our interpretation of a linear extrapolation is based on the 40 year
old philosophy and work of Glauber \cite{glaub} and on the 20 year
old rows on rows model of H\"{u}fner and Knoll \cite{huf} as applied to the
now disassembled Bevalac at LBNL (Lawrence Berkeley National Laboratory).
Nucleons from each nucleus follow straight line trajectories, making
binary collisions with nucleons from the other nucleus.  These collisions
are as in free space.  Inelastic collisions produce mesons; the mesons
are not allowed to collide with each other or with any nucleons.  The
number of binary collisions suffered by any given nucleon depends on
the nucleon cross section and on the geometry of the nuclei.  The details
will be given in later sections.  It is important to know that this
linear extrapolation model which we refer to as LEXUS, for Linear
EXtrapolation of Ultrarelativistic Scattering, has no free parameters.

We will apply LEXUS to published data on S+S and Pb+Pb collisions at the
SPS.  (At this time the quantity of Pb+Pb data available to us
is not as complete as the S+S data.)  We do not attempt to apply LEXUS
to AGS energies.  Those energies are probably too low to accept the
assumption of straightline trajectories as being anywhere near
realistic.

It is important to keep in mind that all LEXUS predictions in this
paper are absolutely normalized.  We have not attempted to tune the
results in any sense.

A conclusion of our paper is that a linear extrapolation of nucleon
collisions is consistent with S+S and Pb+Pb collisions at the SPS in the
sense that it gives a good representation of the baryon rapidity
and transverse momentum distributions, the negative hadron
rapidity and transverse momentum distributions, the average number
of pions, and zero degree energy distributions.  However, it predicts
only about 80\% of the observed number of charged kaons, 50\% of
the number of observed neutral kaons and lambdas,
and 10\% of the number of observed antilambdas, all in
reference to central S+S collisions.  This may suggest where new
physics lies.

In section 2 we formulate the model and solve for the basic building
blocks, the two-particle baryon rapidity distributions.  In section 3
we compute the final, observable baryon rapidity distribution.  In
section 4 we compute the baryon transverse momentum distribution.
In section 5 we compute the average multiplicities of various
produced hadrons.  In section 6 we compute the negative hadron
rapidity distribution.  In section 7 we compute the transverse
momentum distribution of negative hadrons.  In section 8
we compute the zero degree (calorimeter) energy distribution.
Conclusions are drawn in section 9.

\section{Formulation and Solution of the Model}

To formulate the model it is convenient to consider a collision
between two rows of nucleons.  A nucleus-nucleus collision will
be constructed from an ensemble of row-row collisions.
Refer to the nucleons comprising these rows as projectile and target
nucleons.  Let $W_{mn}(y_P,y_T)$ represent the 2-particle rapidity
distribution for the $m$'th projectile nucleon and the $n$'th target
nucleon immediately after their collision.  The single-particle projectile
distribution $W_{mn}^P(y_P)$ is obtained by integrating the 2-particle
distribution over the unobserved target rapidity.  The index $mn$
then refers to the $m$'th projectile nucleon after colliding with
$n$ target nucleons.
\begin{equation}
W_{mn}^P(y_P) = \int dy_T W_{mn}(y_P,y_T)
\end{equation}
Similarly, $W_{mn}^T(y_T)$ is
the single-particle target distribution obtained by integrating over
the unobserved projectile rapidity.  The index $mn$ then refers to
the $n$'th target nucleon after colliding with $m$ projectile nucleons.
These single-particle distributions are normalized to unity.
\begin{equation}
\int dy_P W_{mn}^P(y_P) = \int dy_T W_{mn}^T(y_T) = 1
\end{equation}
Due to the indistinguishability of nucleons the outgoing nucleon
with the larger rapidity is called a projectile and the other
is called a target.

The 2-particle distribution $W_{mn}$ is obtained by the collision
between the $m$'th projectile nucleon, which has suffered $n-1$
previous collisions, with the $n$'th target nucleon, which has
suffered $m-1$ previous collisions.
\begin{equation}
W_{mn}(y_P,y_T) = \int dy_P' dy_T' W_{mn-1}^P(y_P')  W_{m-1n}^T(y_T')
K(y_P' + y_T' \rightarrow y_P + y_T)
\label{evo}
\end{equation}
Here we assume that the process is Markovian with kernel $K$.  This
is not a necessary assumption and could be relaxed.  Doing so would
result in a correlated cascade.  However, it would require experimental
information on the correlation between the two outgoing baryons which
is generally not available.

A basic input is the kernel $K$ which must be taken from experiments
on nucleon-nucleon collisions.  To that end consider a nucleon-nucleon
collision in the LAB frame of reference so that the initial single-particle
distributions are
\begin{eqnarray}
W_{10}^P(y_P') &=& \delta(y_P'-y_0) \nonumber \\
W_{01}^T(y_T') &=& \delta(y_T')
\label{ini}
\end{eqnarray}
where $y_0$ is the beam rapidity.  Substitution into the evolution
equation gives
\begin{equation}
W_{11}(y_P,y_T) = K(y_0+0 \rightarrow y_P+y_T) \, .
\end{equation}
Experiments do not measure the correlated two-nucleon distribution
over all phase space, they only measure the single particle distribution.
\begin{equation}
\frac{dN}{dy}(y) = W_{11}^P(y) + W_{11}^T(y)
\end{equation}
Here the projectile contribution is
\begin{equation}
W_{11}^P(y_P) = \int dy_T W_{11}(y_P,y_T) \, ,
\end{equation}
and similarly for the target contribution.

\begin{figure}[p]
\begin{picture}(0,0)(0,10.3)
\put(0.85,0){$-4$}
\put(2.506,0){$-3$}
\put(4.163,0){$-2$}
\put(5.8188,0){$-1$}
\put(7.675,0){$0$}
\put(9.33125,0){$1$}
\put(10.9875,0){$2$}
\put(12.64375,0){$3$}
\put(14.3,0){$4$}
\put(7.6,-0.5){$y_{\rm cm}$}
\put(0.6,0.35){$0$}
\put(0.5,1.4875){$1$}
\put(0.5,2.675){$2$}
\put(0.5,3.8625){$3$}
\put(0.5,5.05){$4$}
\put(0.5,6.2375){$5$}
\put(0.5,7.425){$6$}
\put(0.5,8.6125){$7$}
\put(0.5,9.8){$8$}
\put(-0.5,5.05){$\displaystyle {dN_p\over dy}$}
\end{picture}
\centerline{\psfig{figure=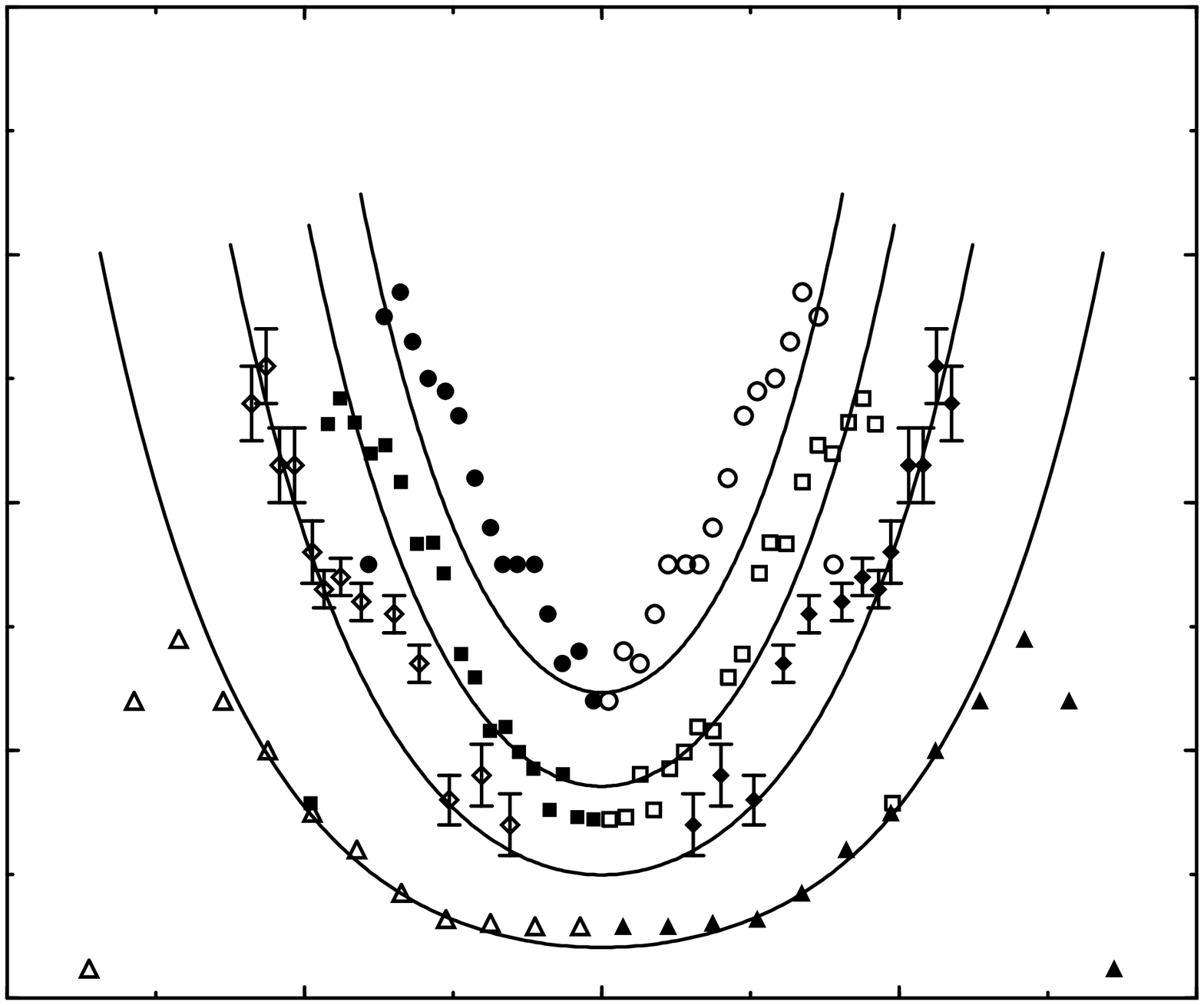,width=14cm,height=10cm}}
\vspace{0.1cm}
\caption{%
Proton rapidity distribution in $p + p \to p + X$ in the center of mass
frame.
From the top:
$p_{\rm lab} = 12, 24, 69$,
and 400 GeV$/c$.  The filled symbols indicate
experimentally measured data points and the empty symbols are the
reflected ones.  The solid lines are
$dN/dy = \lambda \cosh(y)/\sinh(y_0/2)$ with
corresponding maximum rapidity $y_0$.  The data were assembled in
Ref.~\protect\cite{Ole}.
}
\label{fig:pp_cosh}
\end{figure}

It has long been known that, to good approximation, the distribution
of outgoing nucleons in a high energy nucleon-nucleon collision is
flat in longitudinal momentum or a hyperbolic cosine (symmetric about the CM)
in rapidity \cite{Abolins,Whitmore}.  This knowledge does not uniquely
determine the 2-particle kernel $K$ but with the additional, sensible,
requirements that the projectile distribution be forward peaked and that
the simplest mathematical representation be used consistent with the data,
leads to the parametrization
\begin{equation}
K(y_P' + y_T' \rightarrow y_P + y_T) = Q(y_P-y_T',y_P'-y_P,y_P'-y_T')
Q(y_P'-y_T,y_T-y_T',y_P'-y_T')
\label{K}
\end{equation}
where
\begin{equation}
Q(s,t,u) = \lambda \frac{\cosh s}{\sinh t} + (1-\lambda)\delta(u) \, .
\end{equation}
In particular the distribution of outgoing projectile nucleons is
\begin{equation}
W_{11}^P(y) = Q(y,y_0,y_0-y) = \lambda \frac{\cosh y}{\sinh y_0}
+ (1-\lambda)\delta(y_0-y) \, .
\end{equation}
This is the same distribution \cite{mass} as used in the evolution
model proposed by Hwa \cite{Hwa} to describe proton stopping in high energy
proton-nucleus scattering and solved and applied to data by Csernai
and one of the authors \cite{Csernai}.  The distribution is normalized
to unity.  The parameter $\lambda$ is the fraction of nucleon-nucleon
scatterings that result in a hard collision and $1-\lambda$ is the fraction
that are diffractive or elastic.  A recent compilation of data
on $p + p \rightarrow p + X$ in the momentum range 12 to 400 GeV$/c$
leads to $\lambda = 0.6$ \cite{Ole}.  This data and the fit are shown
in Figure~\ref{fig:pp_cosh}.
Unless otherwise stated this is the numerical value used
in the rest of the paper.  Physical observables turn out to be rather
insensitive to small ($\pm 0.1$) variations in $\lambda$.  There is
a small rollover in the data near the projectile and target rapidities
which is not represented by the parametrization.  This has to do with
precisely how one separates hard inelastic and diffractive collisions.
Our results are only as good as the input parametrizations; in the future
it might be worthwhile to treat these components on a finer level.

There is an obvious and useful symmetry between the single-particle
projectile and target distributions.
\begin{equation}
W_{mn}^P(y) = W_{nm}^T(y_0-y)
\end{equation}
These distributions are not independent.  In the present formulation of
the model only single particle observables may be reliably computed.
Hence we only need to compute the $W_{mn}^P(y)$.  For this we need
an evolution equation.  It is obtained by integrating eq. (\ref{evo})
over the target rapidity and using eq. (\ref{K}) for the kernel.
\begin{equation}
W_{mn}^P(y) = \int dy_P dy_T W_{mn-1}^P(y_P) W_{nm-1}^P(y_0-y_T)
Q(y-y_T,y_P-y_T,y-y_P)
\end{equation}
It only remains to solve this Boltzmann-like equation.
This can be accomplished by starting with the initial distribution
eq. (\ref{ini}) and then iterating over all $m$ and $n$.

\begin{figure}[p]
\begin{picture}(0,0)(0,12)
\put(3.05,0){$0$}
\put(4.7,0){$1$}
\put(6.45,0){$2$}
\put(8.2,0){$3$}
\put(9.95,0){$4$}
\put(11.7,0){$5$}
\put(13.45,0){$6$}
\put(8.2,-0.5){$y$}
\put(2.7,0.35){$0$}
\put(2.4,2.23333){$0.1$}
\put(2.4,4.11667){$0.2$}
\put(2.4,6.0){$0.3$}
\put(2.4,7.88333){$0.4$}
\put(2.4,9.76667){$0.5$}
\put(2.4,11.65){$0.6$}
\put(0.9,6.0){$\displaystyle W_{mn}(y)$}
\put(12,8){$W_{3,1}$}
\put(11,6){$W_{3,4}$}
\put(7.3,5.0){$W_{3,7}$}
\put(5.4,6.8){$W_{3,10}$}
\put(4.5,10){$W_{3,13}$}
\end{picture}
\centerline{\psfig{figure=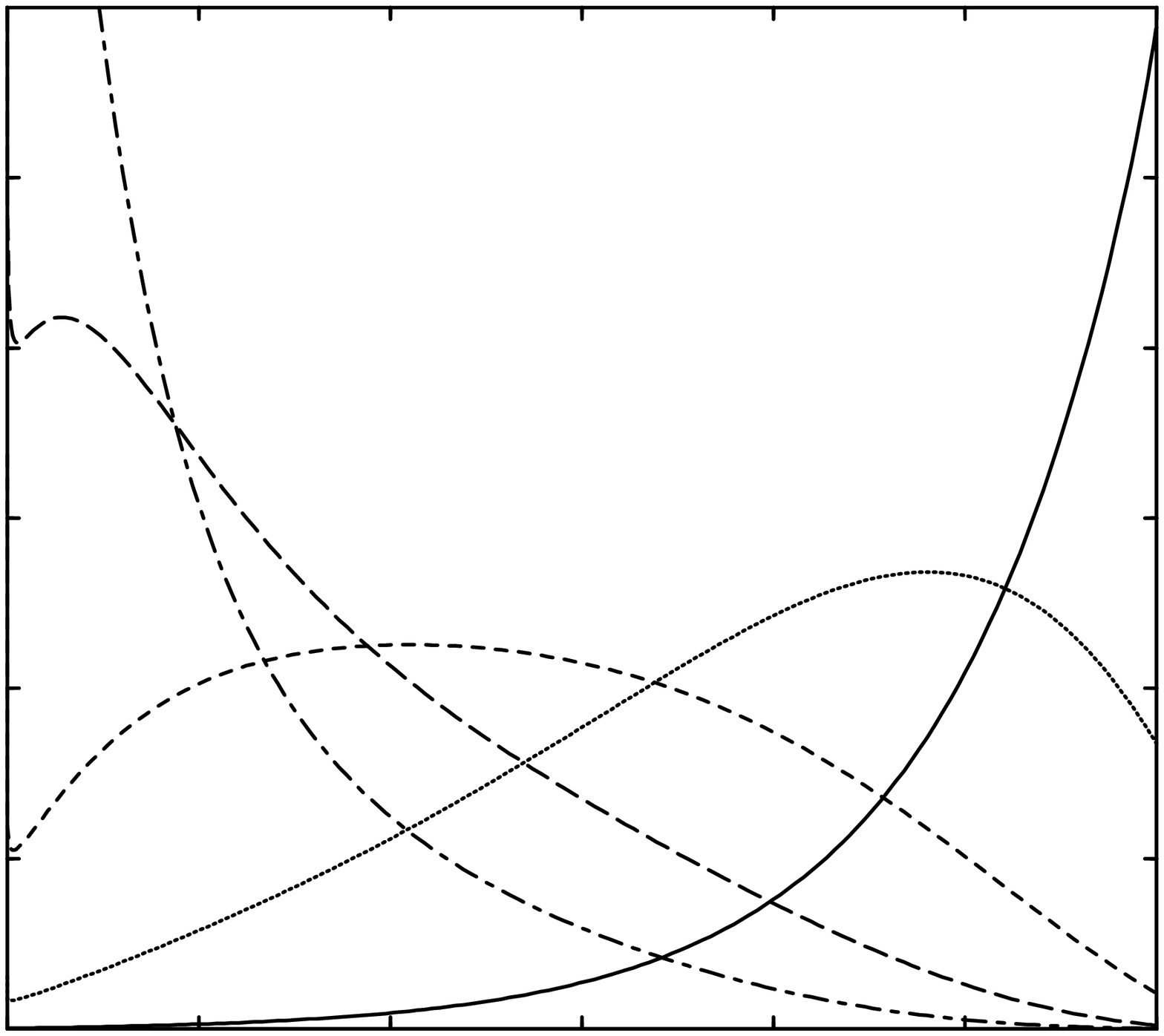,width=12cm,height=12cm}}
\vspace{0.1cm}
\caption{%
Graphs of $W_{3n}(y)$ for $n = \{1,4,7,10,13\}$.  }
\label{fig:w3n}
\end{figure}

A closed form expression can be given for the first nucleon in the
row undergoing an arbitrary number of collisions \cite{Csernai}.
\begin{equation}
W_{1n}^P(y) = \frac{\cosh y}{\sinh y_0} \sum_{k=1}^n
\left( \begin{array}{c} n \\ k \end{array} \right)
\frac{\lambda^k (1-\lambda)^{n-k}}{(k-1)!}
\left[ \ln\left( \frac{\sinh y_0}{\sinh y} \right) \right]^{k-1}
+ (1-\lambda)^n \delta(y_0-y)
\end{equation}
Unfortunately we were not able to find a closed expression for
arbitrary $mn$, so we solved the equations numerically for
$m$ and $n$ up to and including 14.  The solutions have the form
\begin{equation}
W_{mn}^P(y) = \overline{W}_{mn}^P(y) + (1-\lambda)^n \delta(y_0-y)
\end{equation}
where $\overline{W}$ is a continuous function albeit with logarithmic
singularities at $y=0$ and $y=y_0$.  These very soft singularities
are a consequence of the explicit function $Q$ chosen above.  When
comparing with experiment it should be remembered to smooth these
by the experimental resolution of the detectors.  See also
Figure~\ref{fig:pp_cosh} and the previous discussion.
We have constructed numerical tables of $\overline{W}$.  These may be
obtained from the web site of one of the authors \cite{web}.
Some representative examples are plotted in Figure~\ref{fig:w3n}.

\section{Baryon Rapidity Distribution}

In this section we will apply the most obvious output of the model,
the baryon rapidity distribution, to the available experimental data
on nucleus-nucleus collisions.  But first we must describe how
to make a nucleus-nucleus collision out of row on row collisions.
This is standard material for any Glauber-like model so we shall go
over it without too much discussion.

\begin{figure}
\begin{picture}(0,0)(0,12)
\end{picture}
\centerline{\psfig{figure=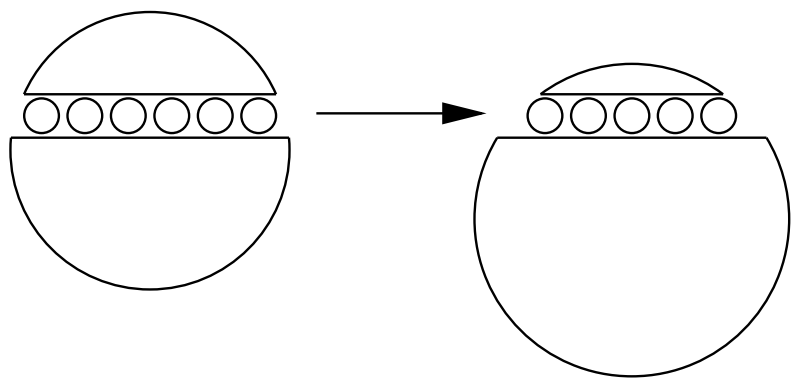}}
\caption{%
A schematic view of a row on row collision.}
\label{fig:row_on_row}
\end{figure}

Consider a collision between a projectile nucleus and a target nucleus
with an impact parameter ${\bf b}$.  We can think of this approximately
as a sum of independent collisions between rows of nucleons as illustrated
in Figure~\ref{fig:row_on_row}.
Two rows will collide when the transverse position
${\bf s}_P$ of the projectile row relative to the projectile nucleus'
center of mass and the transverse position ${\bf s}_T$ of the target
row relative to the target nucleus' center of mass are related by
${\bf s}_T = {\bf b} + {\bf s}_P$.  The average number of nucleons
in each row, with cross sectional area equal to the nucleon-nucleon
cross section, are
\begin{eqnarray}
\nu_P({\bf s}_P) &=& \sigma_{NN}\int dz \rho_P({\bf s}_P,z) \nonumber \\
\nu_T({\bf s}_T) &=& \sigma_{NN}\int dz \rho_T({\bf s}_T,z)
\end{eqnarray}
where $\rho$ is the baryon density and $z$ is the longitudinal coordinate.
Let ${\cal P}^P_{\bar m}({\bf s}_P)$ and ${\cal P}^T_{\bar n}({\bf s}_T)$
denote the probability of having ${\bar m}$ and ${\bar n}$ nucleons
in the projectile and target rows, respectively, at the given impact
parameter.  We defer the actual choice of these probabilities.

There will be fluctuations in the number of nucleons in each row.
Similarly, there will be fluctuations in the number of collisions
suffered by any given nucleon in a row.  Taking these into account
results in the contribution of the projectile nucleons to the final
baryon rapidity distribution as follows.
\begin{equation}
\frac{dN_P}{dy}(y,{\bf b}) = \sum_{{\bar m}=1}^{A_P} \sum_{m=1}^{\bar m}
\sum_{n=0}^{A_T} W_{mn}^P(y) \int \frac{d^2s_P}{\sigma_{NN}}
{\cal P}_n^T({\bf s}_T) {\cal P}_{\bar m}^P({\bf s}_P)
\label{projy}
\end{equation}
There is an analogous expression for the target contribution.  The total
rapidity distribution at the fixed impact parameter is the sum of the
projectile and target contributions.

Now the choice of the ${\cal P}$s must be made.  One candidate is a
Poisson distribution.  Its disadvantage is two-fold: it overestimates
the magnitude of the fluctuations (they are unlimited in a Poisson
but are limited in reality by the number of available nucleons)
and it would require knowledge of the $W_{mn}^P$ for arbitrarily
large values of $m$ and $n$, which is computationally infeasible.
A natural alternative is a binomial; but with what maximum value?
It is physically unreasonable to allow all nucleons in a nucleus to
fluctuate into one row.  The maximum average value of the nucleons
in a given row is about 10 in a large nucleus.  We have chosen the
maximum value to be $N_{\rm max}$ to be 14.  Varying this number by
1 or 2 does not change any distribution by more than a percent.
Explicitly we have chosen
\begin{equation}
{\cal P}_{\bar m}^P({\bf s}_P) =
\left( \begin{array}{c} N_{\rm max} \\ {\bar m}
\end{array} \right) \left[\frac{\nu_P({\bf s}_P)}{N_{\rm max}}
\right]^{\bar m}
\left[ 1-\frac{\nu_P({\bf s}_P)}{N_{\rm max}} \right]^{N_{\rm max}-{\bar m}}
\Theta(N_{\rm max} - {\bar m}) \, . \label{binomial}
\end{equation}
Here $\Theta$ is the step function.  For the nucleon-nucleon
cross section we use a constant value of
4 fm$^2$ which is appropriate for beam energies ranging from tens
to hundreds of GeV.

For the nuclear density distribution in lead we use a Woods-Saxon
\begin{equation}
\rho(r) = \frac{\rho_0}{1 + \exp[(r-b)/a]}
\end{equation}
with parameters $a = 0.546$ fm and $b = 6.62$ fm.
Normalization to $A = 208$ fixes  $\rho_0 = 0.1604$ fm$^{-3}$.
For sulfur we use a three-parameter Gaussian \cite{datatab}
\begin{equation}
\rho(r) = \frac{\rho_0(1+wr^2/b^2)}{1 + \exp[(r^2-b^2)/a^2]}
\end{equation}
with parameters $w = 0.160$, $a = 2.191$ fm, and $b = 2.54$ fm.
Normalization to $A = 32$ fixes $\rho_0 = 0.226$ fm$^{-3}$.

To compare with experimental measurements we must know the trigger
conditions; that is, we must know with what probability any particular
impact parameter is accepted by the detector.  Such a trigger can best
be accommodated by an event generator, for then the output of the theory
can be sent through the experimental filter.  Since our model is
not an event generator, at least in its present form, we can only
attempt to simulate the trigger as best we can.  The procedure we
shall use is to make a sharp impact parameter cutoff.  The total
nucleus-nucleus cross section may be computed following Karol
\cite{Karol}.
\begin{equation}
\sigma^{\rm tot}_{A_PA_T} = \int d^2b \left\{ 1 - \exp\left[
-f({\bf b})\right] \right\}
\end{equation}
Here $f$ is the geometrical overlap function of the two nuclei.
\begin{equation}
f({\bf b}) = \int\frac{d^2s_P}{\sigma_{NN}} \nu_P({\bf s}_P)
\nu_T({\bf s}_T)
\end{equation}
If only those nucleus-nucleus collisions are accepted with an
impact parameter less than or equal to $b_{\rm cut}$ then the
corresponding cross section is
\begin{equation}
\sigma_{A_PA_T}(b_{\rm cut}) = \int d^2b \left\{ 1 - \exp\left[
-f({\bf b})\right] \right\} \Theta(b_{\rm cut}-b) \, .
\end{equation}
Clearly
\begin{equation}
\sigma^{\rm tot}_{A_PA_T} = \lim_{b_{\rm cut} \rightarrow
\infty} \sigma_{A_PA_T}(b_{\rm cut}) \, .
\end{equation}
The impact parameter cutoff can be adjusted to reproduce a given
centrality cut.  For example, if only the 6\% ``most central"
collisions are accepted then $b_{\rm cut}$ is adjusted such that
$\sigma_{A_PA_T}(b_{\rm cut}) = 0.06 \sigma^{\rm tot}_{A_PA_T}$.

\begin{figure}[p]
\begin{picture}(0,0)(0,12)
\put(3.14,0){$0$}
\put(4.7,0){$1$}
\put(6.45,0){$2$}
\put(8.2,0){$3$}
\put(9.95,0){$4$}
\put(11.7,0){$5$}
\put(13.45,0){$6$}
\put(8.2,-0.5){$y$}
\put(2.7,0.35){$0$}
\put(2.7,1.7625){$1$}
\put(2.7,3.175){$2$}
\put(2.7,4.5875){$3$}
\put(2.7,6.0){$4$}
\put(2.7,7.4125){$5$}
\put(2.7,8.825){$6$}
\put(2.7,10.2375){$7$}
\put(2.7,11.65){$8$}
\put(1.2,6.0){$\displaystyle {dN_p\over dy}$}
\end{picture}
\centerline{\psfig{figure=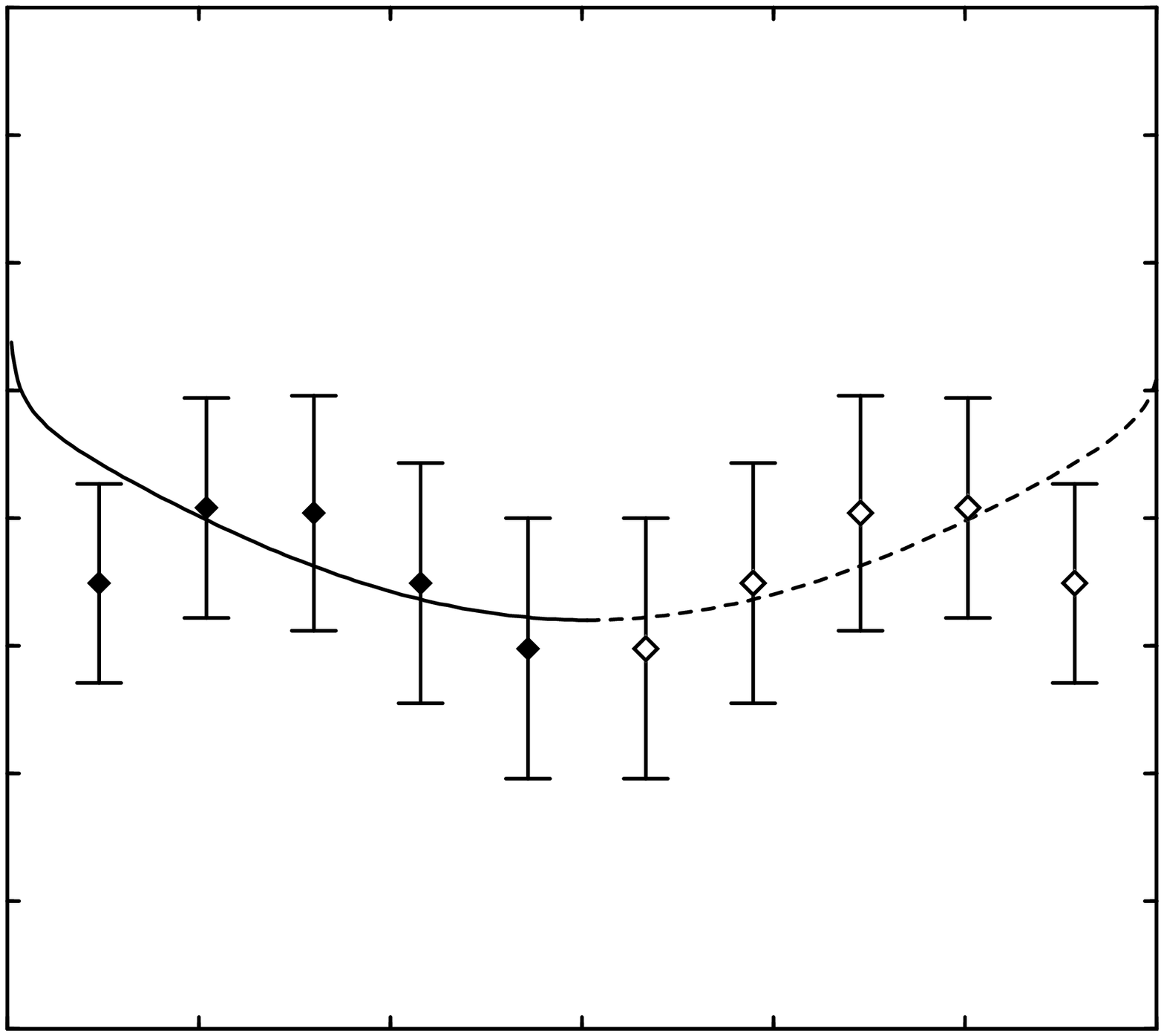,width=12cm,height=12cm}}
\vspace{0.1cm}
\caption{%
Proton rapidity distribution for a S+S collision
at 200 GeV$/c$ with a centrality of 2\%.
The solid line represents our calculation.
Filled diamonds are data from NA35.  Open diamonds
and the dashed line are the reflection of the left half. }
\label{fig:pdndy_NA35}
\end{figure}

Now we compare with experiment.  Figure~\ref{fig:pdndy_NA35}
shows the proton rapidity
distribution measured by NA35 \cite{NA35dist} for the 2\% most
central collisions of S+S at a beam energy of 200 GeV$/c$ per nucleon.
The solid symbols are the actual measurements and the open symbols
are those obtained by reflection about midrapidity.  Except possibly
for the nuclear fragmentation regions, that is, within $\pm1$ unit
of rapidity of beam and target, the data is represented very well
by LEXUS.

\begin{figure}[p]
\begin{picture}(0,0)(0,12)
\put(3.05,0){$0$}
\put(4.7,0){$1$}
\put(6.45,0){$2$}
\put(8.2,0){$3$}
\put(9.95,0){$4$}
\put(11.7,0){$5$}
\put(13.45,0){$6$}
\put(8.2,-0.5){$y$}
\put(2.7,0.35){$0$}
\put(2.5,2.61){$10$}
\put(2.5,4.87){$20$}
\put(2.5,7.13){$30$}
\put(2.5,9.39){$40$}
\put(2.5,11.65){$50$}
\put(1.2,6.0){$\displaystyle {dN_p\over dy}$}
\end{picture}
\centerline{\psfig{figure=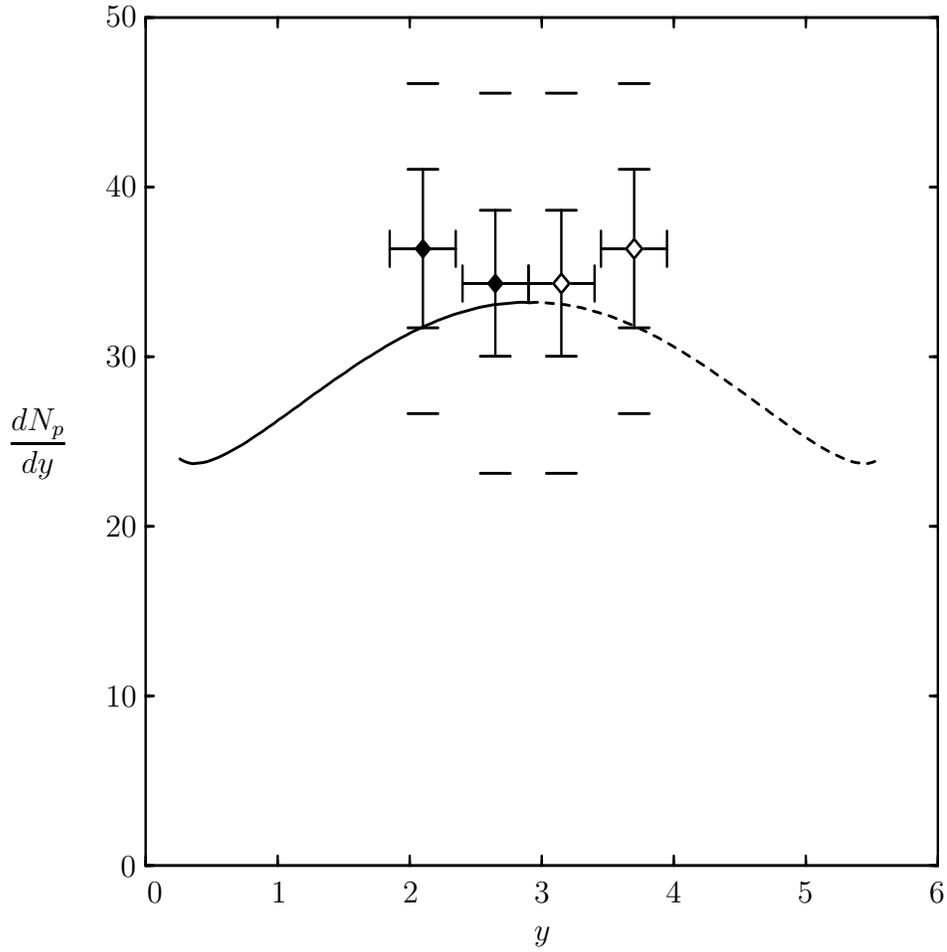,width=12cm,height=12cm}}
\vspace{0.1cm}
\caption{%
Proton rapidity distribution for a Pb+Pb collision at 158 GeV$/c$ with a
centrality of 6.4\%.
The solid line is our result.  The data points from NA44 are marked by
filled diamonds.  The open diamonds and the dashed lines are the
reflection of the left half.  Error bars on the NA44 data represent
statistical errors, the limiting short bars represent systematic
errors.}
\label{fig:pdndy_NA44}
\end{figure}

The first measurements of the proton rapidity distribution in
Pb+Pb collisions at 158 GeV$/c$ per nucleon have just recently been
published by NA44 \cite{NA44prot}.  They are shown
in Figure~\ref{fig:pdndy_NA44} and
represent the 6.4\% most central collisions.
The systematic plus statistical uncertainties together
are very large, and there are only two measured points.
To compare with the proton distributions in Pb+Pb we should
point out an uncertainty in our current application of LEXUS.
We have not distinguished between outgoing protons and neutrons
but have only counted baryons.  For collisions of charge
asymmetric nuclei at high energy it is reasonable to expect that
nearly 1/2 of the outgoing baryons will be protons due to the
preference to convert more neutrons into protons than vice versa.
This simply follows from phase space and entropy.  Accepting that,
the LEXUS predictions are in very good agreement with the data.

\begin{figure}[p]
\begin{picture}(0,0)(0,12)
\put(3.05,0){$0$}
\put(4.7,0){$1$}
\put(6.45,0){$2$}
\put(8.2,0){$3$}
\put(9.95,0){$4$}
\put(11.7,0){$5$}
\put(13.45,0){$6$}
\put(8.2,-0.5){$y$}
\put(2.7,0.35){$0$}
\put(2.5,2.61){$10$}
\put(2.5,4.87){$20$}
\put(2.5,7.13){$30$}
\put(2.5,9.39){$40$}
\put(2.5,11.65){$50$}
\put(1.2,6.0){$\displaystyle {dN_p\over dy}$}
\end{picture}
\centerline{\psfig{figure=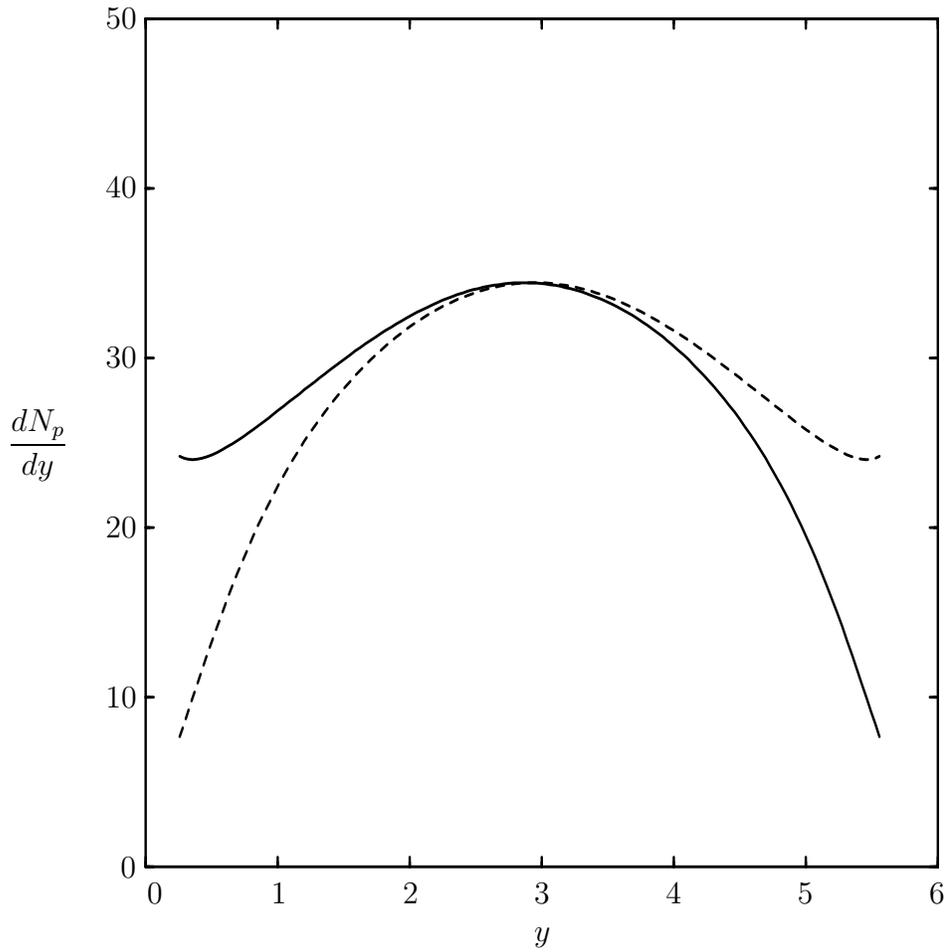,width=12cm,height=12cm}}
\vspace{0.1cm}
\caption{%
Proton rapidity distribution for a Pb+Pb collision
at 158 GeV$/c$ with 5\% centrality.
The solid line is our result, the dashed line is the reflection
about the mid-rapidity.
The asymmetry is caused by the high rapidity particle loss
to the zero degree calorimeter.}
\label{fig:pdndy_NA49}
\end{figure}

So far NA49 has not published its measurements of the baryon
rapidity distribution in Pb+Pb collisions.  However, we would
like to point out a curious feature of the symmetry between
forward and backward going particles in the center of mass frame.
In an ideal measurement it would be symmetric in collisions
of identical nuclei when averaged over a sufficient number of
collisions.  But if there is a zero degree calorimeter used
to select on central collisions then any particles going into
that calorimeter may be considered as lost and not counted in
the rapidity distribution.  This induces a forward/backward
asymmetry.  The effect is shown in LEXUS in Figure~\ref{fig:pdndy_NA49}
(details will be discussed in section 8).  The upper symmetric
curve represents the proton distribution without a zero degree
calorimeter.  The solid curve presents the depletion of protons
in the forward direction.  It is clear that different results are
obtained if one measures only for $y \ge 3$ and reflects about
$y = 3$ than if one measures only for $y \le 3$ and then reflects.

\section{Baryon Transverse Momentum Distribution}

Even though LEXUS assumes straightline trajectories it is still
possible to get an enhancement of the baryon transverse momentum
compared to nucleon-nucleon collisions.  Every time a collision
occurs the baryons can get a transverse kick.  As long as they
continue to travel with high velocity this will not negate the
assumption of straightline trajectories.  The path of the baryon
may be thought of as a random walk in transverse momentum space.
The average transverse momentum-squared of a baryon is related to
the average number $k$ of collisions it has suffered according to
\begin{equation}
\langle p_T^2 \rangle_k = k \langle p_T^2 \rangle_{NN}
\end{equation}
where $\langle p_T^2 \rangle_{NN}$ is the average in a
nucleon-nucleon collision.  This quantity is approximately
beam energy independent for the beam energies of interest
to us \cite{Eisen,Zab}.  We will use $\langle p_T^2 \rangle_{NN}
= 0.282\ (\hbox{GeV}/c)^2$.

We will assume that the rapidity and transverse momentum distributions
factorize in the elementary nucleon-nucleon collisions, which is close
to fact except near the edges of phase space.  The transverse momentum
distribution is well represented by the thermal form
\begin{equation}
\frac{dN}{p_T dp_T} = {\cal N}_1 m_T K_1(m_T/T_1)
\label{therm}
\end{equation}
where $m_T = \sqrt{m_N^2 + p_T^2}$ is the proton transverse mass
and ${\cal N}_1$ normalizes the distribution to one.
Taking $\langle p_T^2 \rangle_{NN} = 0.282\ (\hbox{GeV}/c)^2$ converts
into $T_1 = 113$ MeV.  After suffering $k$ collisions the baryon
transverse momentum distribution becomes broader as reflected in
a higher temperature $T_k$ \cite{temp} which is determined by
\begin{equation}
{\cal N}_k \int_0^{\infty} dp_T\,p_T^3\, m_T\, K_1(m_T/T_k)
= k \langle p_T^2\rangle_{NN} \, .
\end{equation}
The transverse momentum distribution of projectile baryons can
now be computed by averaging over the number of collisions.
For $y \ne y_0$, $p_T \ne 0$ it is
\begin{eqnarray}
\frac{d^2N^P}{p_Tdp_Tdy} &=& \sum_{{\bar m}=1}^{A_P} \sum_{m=1}^{\bar m}
\sum_{n=1}^{A_T} \frac{\overline{W}_{mn}^P(y)}{1-(1-\lambda)^n}
\sum_{k=1}^n
\left( {n\atop k} \right)\,
\lambda^k (1-\lambda)^{n-k}
{\cal N}_k m_T K_1(m_T/T_k)
\nonumber \\ &\times& \int \frac{d^2b}
{\sigma_{A_PA_T}(b_{\rm cut})} \Theta(b_{\rm cut} - b)
\int \frac{d^2s_P}{\sigma_{NN}}
{\cal P}_n^T({\bf s}_T) {\cal P}_{\bar m}^P({\bf s}_P) \, .
\label{protpt}
\end{eqnarray}
The reason for the sum over $k$ at fixed $n$ is that the probability
of any given scattering to be considered hard is $\lambda$,
which is binomially distributed.  The $k = 0$ term just results in
$p_T = 0$.  There is an analogous expression for target nucleons.

\begin{figure}[p]
\begin{picture}(0,0)(0.1,12)
\put(3.15,0){$0$}
\put(4.0,0){$0.2$}
\put(5.05,0){$0.4$}
\put(6.1,0){$0.6$}
\put(7.15,0){$0.8$}
\put(8.2,0){$1.0$}
\put(9.25,0){$1.2$}
\put(10.3,0){$1.4$}
\put(11.35,0){$1.6$}
\put(12.4,0){$1.8$}
\put(13.45,0){$2.0$}
\put(7.7,-0.5){$p_T\ (\hbox{GeV}/c)$}
\put(2.35,0.35){$0.01$}
\put(2.5,3.175){$0.1$}
\put(2.5,6.0){$1$}
\put(2.5,8.825){$10$}
\put(2.4,11.65){$100$}
\put(0.8,4.5)
{\begin{sideways}
$\displaystyle {1\over p_T}{dN_p\over dp_T}\
\left((\hbox{GeV}/c)^{-2}\right)$
\end{sideways}}
\end{picture}
\centerline{\psfig{figure=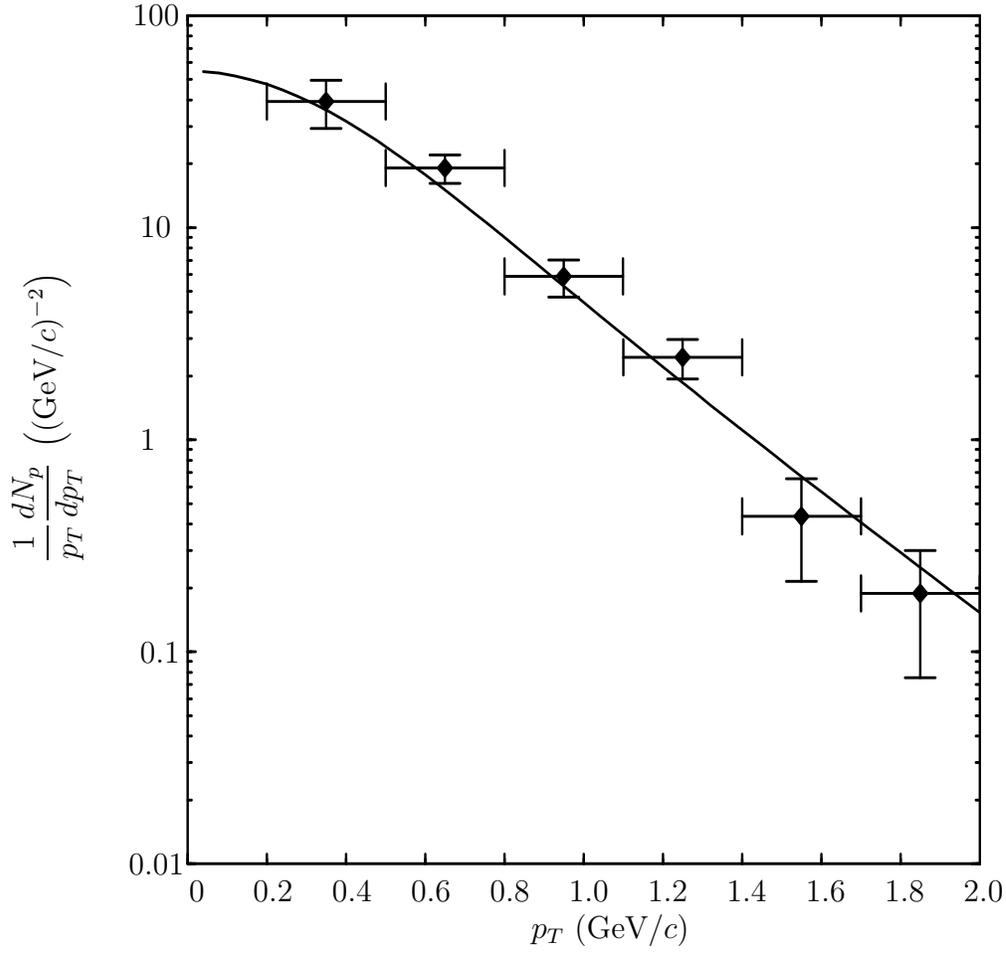,width=12cm,height=12cm}}
\vspace{0.1cm}
\caption{%
Proton transverse momentum distribution for a S+S collision
at 200 GeV$/c$ with a centrality of 2\%.
The solid line represents our calculation.
Data are from NA35.  Rapidity range is $0.2 < y < 3.0$.}
\label{fig:ppt_NA35}
\end{figure}

Now we compare with experiment.  The transverse momentum distribution
in the 2\% most central S+S collisions at 200 GeV$/c$ per nucleon has
been measured by NA35 \cite{NA35dist}.
The data span the rapidity range $0.2 < y < 3.0$
and are shown in Figure~\ref{fig:ppt_NA35}.
The shape and the absolute normalization
are in very good agreement with the results of LEXUS.

\begin{figure}[p]
\begin{picture}(0,0)(0.1,12)
\put(3.15,0){$0$}
\put(5.05,0){$0.2$}
\put(7.15,0){$0.4$}
\put(9.25,0){$0.6$}
\put(11.35,0){$0.8$}
\put(13.45,0){$1.0$}
\put(7.7,-0.5){$m_T - m\ (\hbox{GeV}/c^2)$}
\put(2.35,0.35){$0.01$}
\put(2.5,3.175){$0.1$}
\put(2.5,6.0){$1$}
\put(2.5,8.825){$10$}
\put(2.4,11.65){$100$}
\put(0.8,4.0){
\begin{sideways}
$\displaystyle {1\over 2\pi m_T}{d^2 N_p\over dm_T dy}\
\left( (\hbox{GeV}/c^2)^{-2} \right)$
\end{sideways}}
\end{picture}
\centerline{\psfig{figure=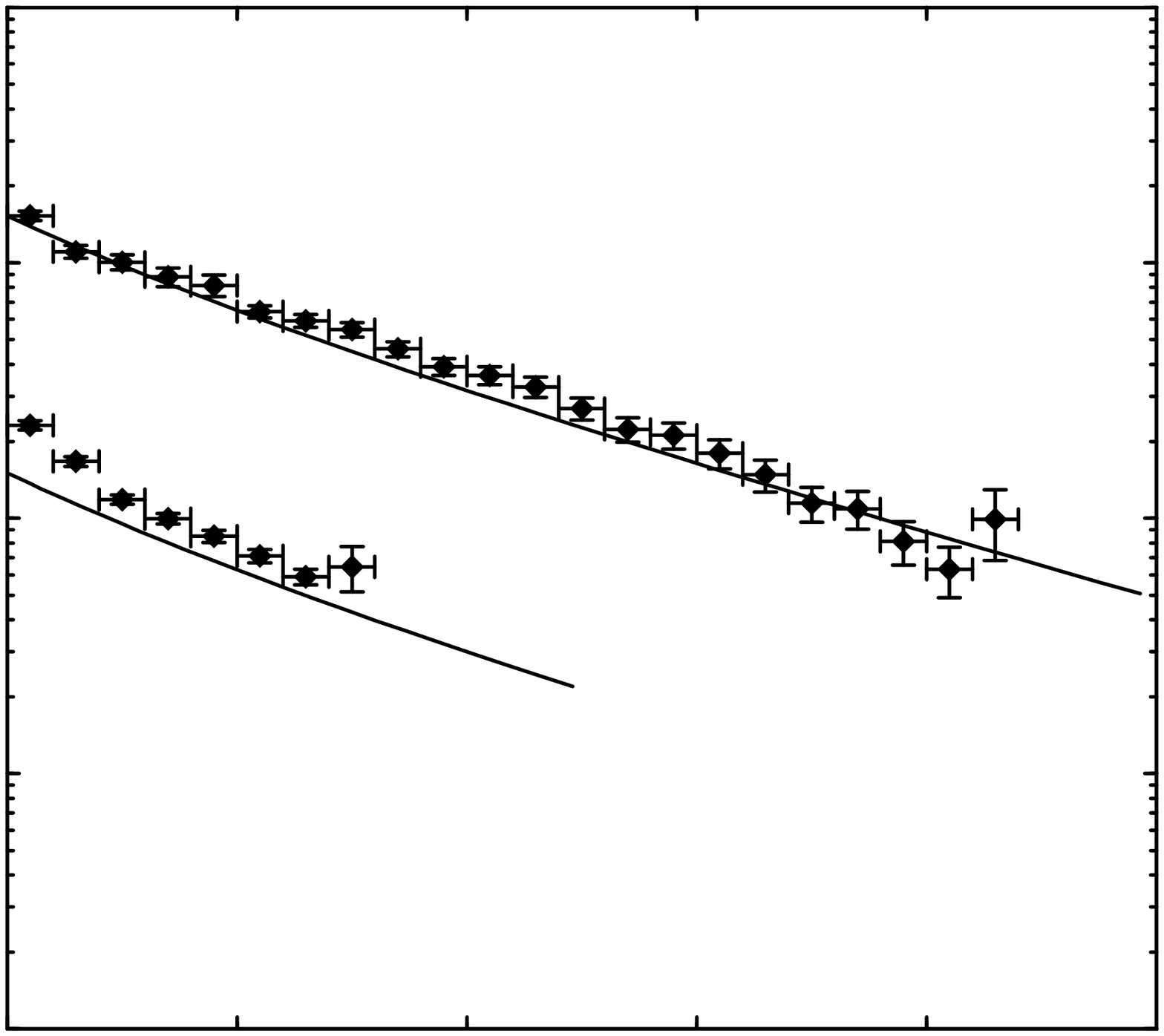,width=12cm,height=12cm}}
\vspace{0.1cm}
\caption{%
Proton transverse momentum distribution for a Pb+Pb collision
at 158 GeV$/c$ with a centrality of 6.4\%.
Filled diamonds are NA44 data and the solid curve is our result.
The upper curve is for $y = 2.65$.
The lower curve is for $y = 2.10$ scaled by a factor of $1/10$.}
\label{fig:ppt_NA44}
\end{figure}

The transverse mass distribution for protons from
the 6.4\% most central Pb+Pb collisions
at 158 GeV$/c$ per nucleon have been measured by NA44 \cite{NA44prot}.
The measurements were performed at $y = 2.10$ and at $y = 2.65$.
They are shown in comparison with LEXUS
in Figure~\ref{fig:ppt_NA44}.  The higher
rapidity data is again in very good agreement with LEXUS; the lower
rapidity data is in less good agreement.  The most likely explanation
is that we have assumed that elastically or diffractively scattered
nucleons acquire zero transverse momentum.  Allowing for it would
increase the distribution at small $p_T$ nearer the projectile
and target rapidities.  This is a topic for future investigation.

\section{Average Multiplicities of Produced Hadrons}

The production of secondary hadrons, such as pions and kaons, or
the conversion of the incident nucleons to hyperons, carries
additional information about the collision dynamics, in particular
the entropy.  In LEXUS we assume that mesons are created in the
collisions of the cascading baryons as in free space.  Once created,
the mesons are assumed to freely disperse: they no longer participate
in the nuclear collision.  The incident nucleons may very well change
their character as they cascade: protons may convert to neutrons and
vice versa, or they may be excited into various $\Delta$, $N^*$, or
hyperon states.  In this section we will compute the average multiplicities
of: charged hadrons $h^-$; kaons $K^+$, $K^-$, $K^0_S$; hyperons
$\Lambda$ and $\overline{\Lambda}$.  We
could compute the full multiplicity distributions, but in this paper
we are content to get the average multiplicities.

Let $\langle X(s) \rangle_{NN}$ represent the average number of
mesons of type $X$ produced in a nucleon-nucleon collision at
center-of-mass energy $\sqrt{s}$.  The average number of such
mesons produced in a collision between the $m$'th projectile nucleon
and the $n$'th target nucleon is
\begin{equation}
\langle X_{mn} \rangle = \lambda \int dy_P dy_T
W_{mn-1}^P(y_P) W_{m-1n}^T(y_T) \langle X(y_P-y_T) \rangle_{NN} \, ,
\label{number1}
\end{equation}
where $\sqrt{s} = 2m_N\cosh\left[(y_P-y_T)/2\right]$.  The total number
produced in a given row-row collision is obtained by summing over
all nucleon-nucleon collisions.
\begin{equation}
\langle X({\bf s}_P,{\bf s}_T) \rangle = \sum_{\bar{m}=1}^{A_P}
\sum_{\bar{n}=1}^{A_T}
{\cal P}_{\bar n}^T({\bf s}_T) {\cal P}_{\bar m}^P({\bf s}_P)
\sum_{m=1}^{\bar{m}} \sum_{n=1}^{\bar{n}}
\langle X_{mn} \rangle
\label{number2}
\end{equation}
Finally we need to sum over all rows and over all allowed impact
parameters.
\begin{equation}
\langle X(b < b_{\rm cut}) \rangle = \int \frac{d^2b}
{\sigma_{A_PA_T}(b_{\rm cut})} \Theta(b_{\rm cut} - b)
\int \frac{d^2s_P}{\sigma_{NN}} \langle X({\bf s}_P,{\bf s}_T) \rangle
\label{number3}
\end{equation}
This can be written as a trace over the product of two matrices.

Let us first consider the production of negatively charged hadrons.
Ga\'zdzicki has shown \cite{comp} that the average multiplicity of
charged hadrons in isospin averaged nucleon-nucleon collisions
for laboratory momenta ranging from 2 to 400 GeV$/c$ is
fit to within about 6\% by the simple parametrization
\begin{equation}
\langle h^- \rangle_{NN} = 0.784 F_{NN\pi}(s)
\end{equation}
where $F_{NN\pi}$ is the Fermi variable modified for the pion
production threshold
\begin{equation}
F_{NN\pi}(s) = \frac{(\sqrt{s} - 2m_N -m_{\pi})^{3/4}}{s^{1/8}}
\end{equation}
and where $\sqrt{s}$ is given in GeV.  There are two caveats to using
this parametrization in LEXUS.  The first is that for light nuclei such
as oxygen or sulfur the isospin averaging is alright, but for heavier
nuclei such as gold or lead it is not.  However, after one or two
collisions protons are more likely to transform into neutrons than
vice versa because of phase space or entropy.  Then isospin averaging
becomes a better approximation.  The second is that occasionally a nucleon
can convert into a hyperon.  There is essentially no experimental
information on the charged hadron multiplicity in a hyperon-nucleon
collision, for good reason.  Therefore, for lack of any other information,
we shall continue to use Ga\'zdzicki's parametrization.

There is somewhat less experimental information on the multiplicity of
kaons produced in nucleon-nucleon collisions.  These data have
been compiled by Ga\'zdzicki and R\"ohrich \cite{scomp} and shown to be
rather well-defined functions of the Fermi variable.  The
parametrizations we shall adopt are given below.
\begin{eqnarray}
\langle K^+ \rangle_{NN} &=& {a F_{NNK^+}^2 \over b + (F_{NNK^+}-c)^2}
\nonumber \\
\langle K^- \rangle_{NN} &=& {a F_{NNK^-}^3 \over b + (F_{NNK^-}-c)^2}
\nonumber \\
\langle K^0_S \rangle_{NN} &=& {a F_{NNK^0_S}^3 \over b + (F_{NNK^0_S}-c)^2}
\label{eq:fit1}
\end{eqnarray}
The parameters $a,b,c$ are displayed in Table 1.

The experimental data on $\Lambda$ and $\overline{\Lambda}$ production
was also analyzed by Ga\'zdzicki and R\"ohrich.  We have constructed
the following parametrizations for use in LEXUS.
\begin{eqnarray}
\langle \Lambda \rangle_{NN} &=& {a F_{NN\Lambda}^2 \over b +
(F_{NN\Lambda}-c)^2} \nonumber \\
\langle \overline{\Lambda} \rangle_{NN} &=&
{a F_{NN\bar{\Lambda}}^3 \over b + (F_{NN\bar{\Lambda}}-c)^2}
\label{eq:fit2}
\end{eqnarray}
These parameters $a, b, c$ are also listed in Table 1.

\begin{figure}[p]
\begin{picture}(0,0)(0.1,12)
\put(3.15,0){$0$}
\put(4.0,0){$0.5$}
\put(5.25,0){$1$}
\put(6.1,0){$1.5$}
\put(7.35,0){$2$}
\put(8.2,0){$2.5$}
\put(9.45,0){$3$}
\put(10.3,0){$3.5$}
\put(11.55,0){$4$}
\put(12.4,0){$4.5$}
\put(13.65,0){$5$}
\put(7.2,-0.5){$F_{NN\Lambda}\ \left(\sqrt{\hbox{GeV}}\right)$}
\put(2.8,0.35){$0$}
\put(2.3,1.7625){$0.02$}
\put(2.3,3.175){$0.04$}
\put(2.3,4.5875){$0.06$}
\put(2.3,6.0){$0.08$}
\put(2.3,7.4125){$0.1$}
\put(2.3,8.825){$0.12$}
\put(2.3,10.2375){$0.14$}
\put(2.3,11.65){$0.16$}
\put(1.5,6.0){$\displaystyle N_\Lambda$}
\end{picture}
\centerline{\psfig{figure=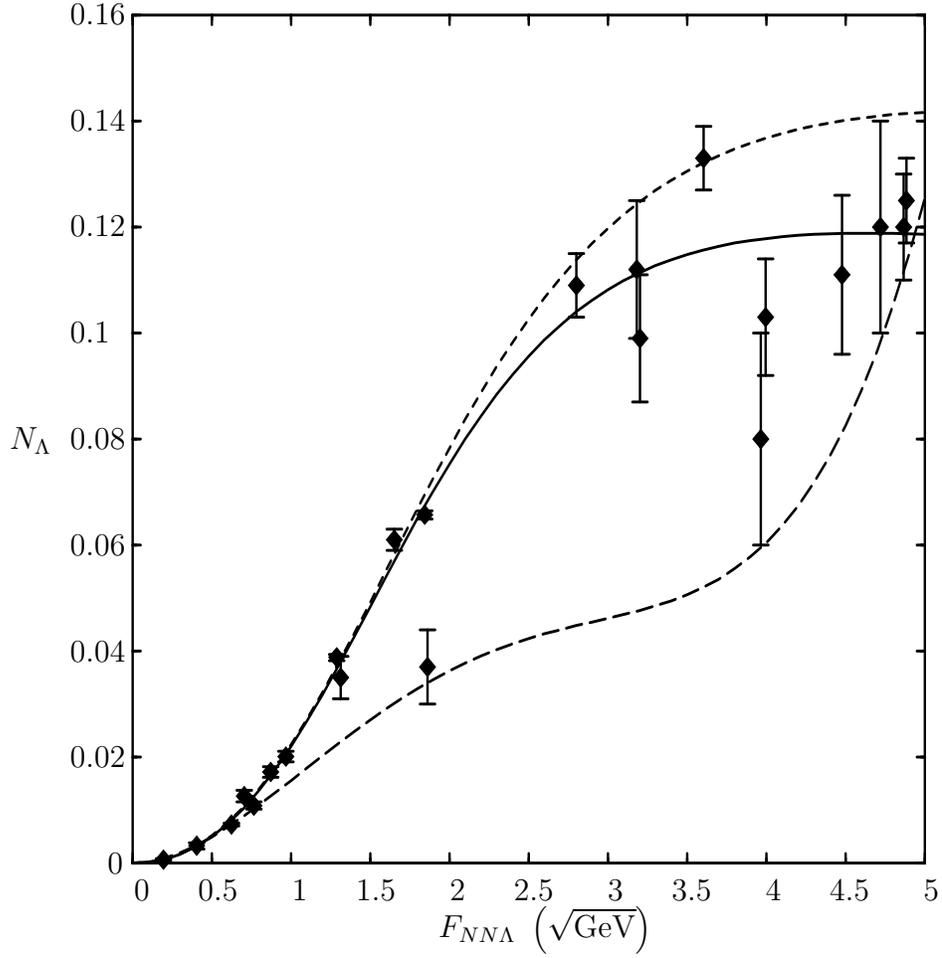,width=12cm,height=12cm}}
\vspace{0.1cm}
\caption{%
Data and fits for $\Lambda$ hyperon production in $p + p$ collisions.
The middle line is the weighted fit for the whole data, the upper
and the lower lines represent maximum and the minimum envelopes.
Experimental data are taken from Ref.~\protect\cite{comp}.}
\label{fig:lambda_pp}
\end{figure}

In all cases above the $F_{NNX}$ are defined as
\begin{equation}
F_{NNX} = {(\sqrt{s} - M_{X})^{3/4} \over s^{1/8}} \, ,
\end{equation}
where the threshold energies are:
$M_X = M_N + M_\Lambda + M_{K^+}$ for $X = \Lambda$, $K^+$ and $K^0_S$;
$M_X = 2 M_N + M_{K^+} + M_{K^-}$ for $K^{-}$; and
$M_X = 2 M_N + 2 M_\Lambda$ for $\bar\Lambda$.
We have taken into account the uncertainties in the production
cross sections in the elementary nucleon-nucleon collisions as
illustrated in Figure~\ref{fig:lambda_pp} for the $\Lambda$ hyperon.
In addition
to making a best fit to the data, upper and lower envelopes are
constructed which roughly pass through the upper and lower error bars,
respectively.  (The parameters for these envelopes are not tabulated
here.)

Calculated results for S+S collisions are shown in Table 2 and compared
with measurements of NA35 \cite{NA35mult1,NA35mult2}
for 2\% centrality.  The LEXUS
results are given for 0\%, 2\% and 4\% most central collisions
based on impact parameter.  These represent the variation in
computed abundances with small variations in impact parameter
away from $b = 0$.  The experimental data is based on a definition
of centrality as those collisions with the highest value of
transverse energy.  This does not exactly correspond to a sharp
range of impact parameter because of fluctuations.  A better
comparison would involve impact parameter smearing.  That is
outside the scope of this paper.

The average number of negatively charged hadrons predicted by LEXUS
is in very good agreement with the measurements.  The average number
of positive or negative kaons obtained from LEXUS is about 80\% of
that observed.  This is a two standard deviation effect when account
is made of the uncertainty in the production rate in nucleon-nucleon
collisions.  The number of short-lived neutral kaons is a factor of
2 too small, as is the number of lambdas.  The number of antilambdas
is nearly an order of magnitude too small compared to experiment.
This almost certainly indicates a failure of the model, most likely
as a result of the neglect of multiple scattering of produced mesons.

Table 3 is a prediction of LEXUS for the 5\% most central collisions
of Pb+Pb at 158 GeV$/c$ per nucleon.  No data has yet been published
for these abundances.

\section{Rapidity Distribution of Secondaries}

It is important to know where the produced particles emerge in
rapidity space.  Several types of detectors are able to measure the
negative charged hadron rapidity distribution $d\langle h^-\rangle/dy$.

In nucleon-nucleon collisions the charged particle rapidity
distribution is approximately Gaussian
\begin{equation}
\frac{d\langle h^-\rangle_{NN}}{dy} = \frac{\langle h^- \rangle_{NN}}
{\sqrt{2\pi} \sigma_L(y_{\rm rel})}
\exp\left[-\left(y-y_{\rm cm}\right)^2/2\sigma_L^2(y_{\rm rel})\right]
\label{density}
\end{equation}
with a width given by the Landau model
\begin{equation}
\sigma_L^2(y_{\rm rel}) = \frac{8}{3}\frac{c_0^2}{1-c_0^4}
\ln \left(\frac{\sqrt{s}}{2m_N}\right)
\end{equation}
where $y_{\rm rel} = y_P-y_T$ is the relative rapidity of
projectile and target and $y_{\rm cm} = \frac{1}{2}\left(y_P+y_T\right)$
is the rapidity of the center-of-mass.  $c_0$ is the speed of sound
of the produced matter.  Not surprisingly $c_0^2 \approx 1/3$ provides
a good fit to nucleon-nucleon data \cite{Landau}.  We have used
a very slightly smaller value of 0.32 corresponding to a free gas
of massive pions at a temperature of 140 to 160 MeV.  The distinction
between 1/3 and 0.32 is irrelevant for the purposes of this paper.

It is straightforward to compute the distribution $d\langle h^-\rangle/dy$
in nucleus-nucleus collisions.  All we need to do is replace the
quantity $\langle X \rangle_{NN}$ with expression
(\ref{density}) in eqs. (\ref{number1})-(\ref{number3}).

\begin{figure}[p]
\begin{picture}(0,0)(0,12)
\put(3.05,0){$0$}
\put(4.817,0){$1$}
\put(6.583,0){$2$}
\put(8.35,0){$3$}
\put(10.12,0){$4$}
\put(11.88,0){$5$}
\put(13.65,0){$6$}
\put(8.2,-0.5){$y$}
\put(2.7,0.35){$0$}
\put(2.4,2.23333){$5$}
\put(2.4,4.11667){$10$}
\put(2.4,6.0){$15$}
\put(2.4,7.88333){$20$}
\put(2.4,9.76667){$25$}
\put(2.4,11.65){$30$}
\put(0.7,6.0){$\displaystyle {dN_{-}\over dy}$}
\end{picture}
\centerline{\psfig{figure=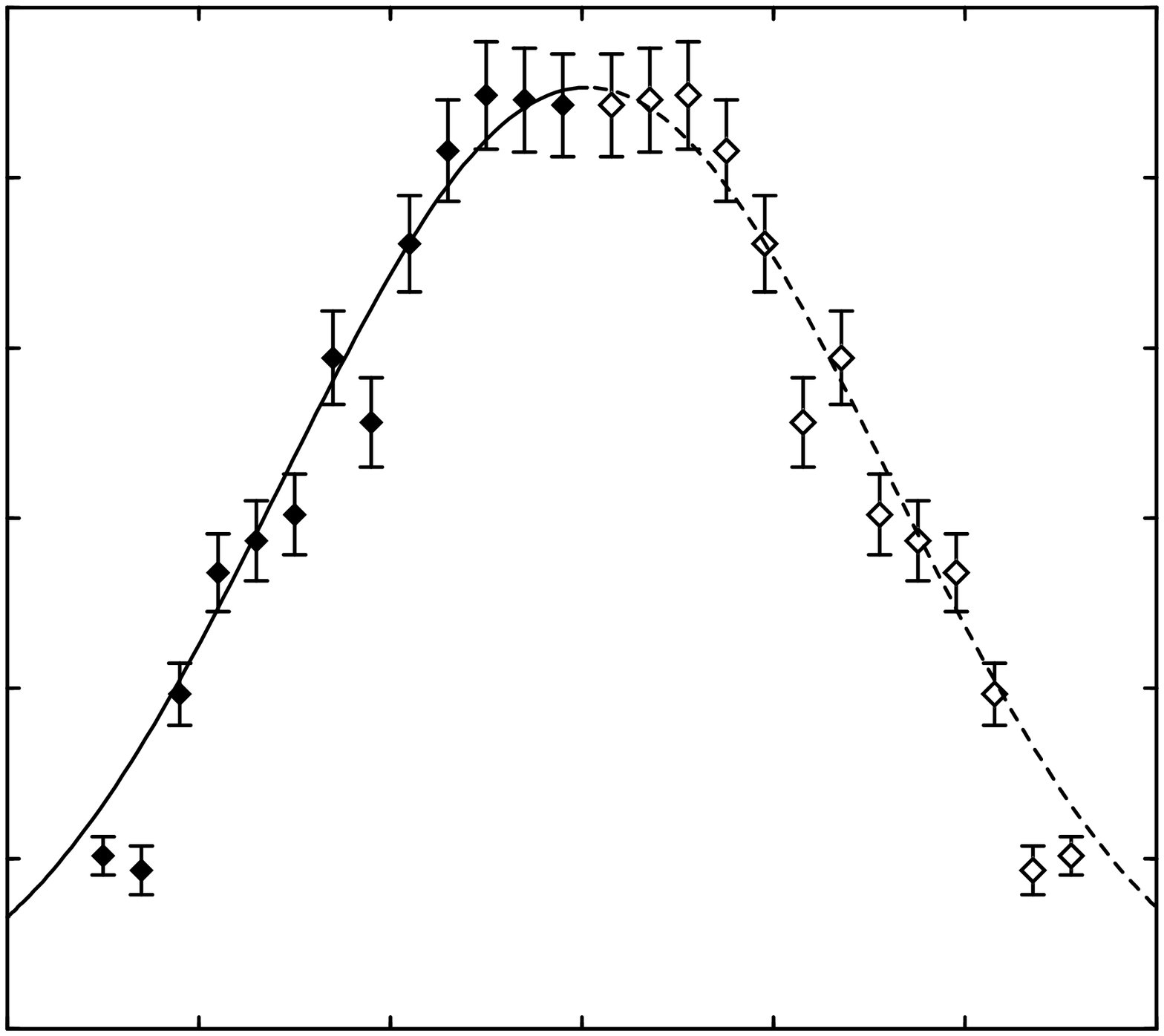,width=12cm,height=12cm}}
\vspace{0.1cm}
\caption{%
Rapidity distribution of $h^-$ in 200 GeV$/c$ S+S collisions with a
centrality of 2\%.
The solid line represents our result.
Filled diamonds are data from NA35.
Open diamonds and the dashed line are the reflection of the left half. }
\label{fig:hmdndy_NA35}
\end{figure}

\begin{figure}[p]
\begin{picture}(0,0)(0,12)
\put(3.05,0){$0$}
\put(4.7,0){$1$}
\put(6.45,0){$2$}
\put(8.2,0){$3$}
\put(9.95,0){$4$}
\put(11.7,0){$5$}
\put(13.45,0){$6$}
\put(8.2,-0.5){$y$}
\put(2.6,0.35){$0$}
\put(2.4,2.61){$50$}
\put(2.3,4.87){$100$}
\put(2.3,7.13){$150$}
\put(2.3,9.39){$200$}
\put(2.3,11.65){$250$}
\put(1.0,6.0){$\displaystyle {dN_{-}\over dy}$}
\end{picture}
\centerline{\psfig{figure=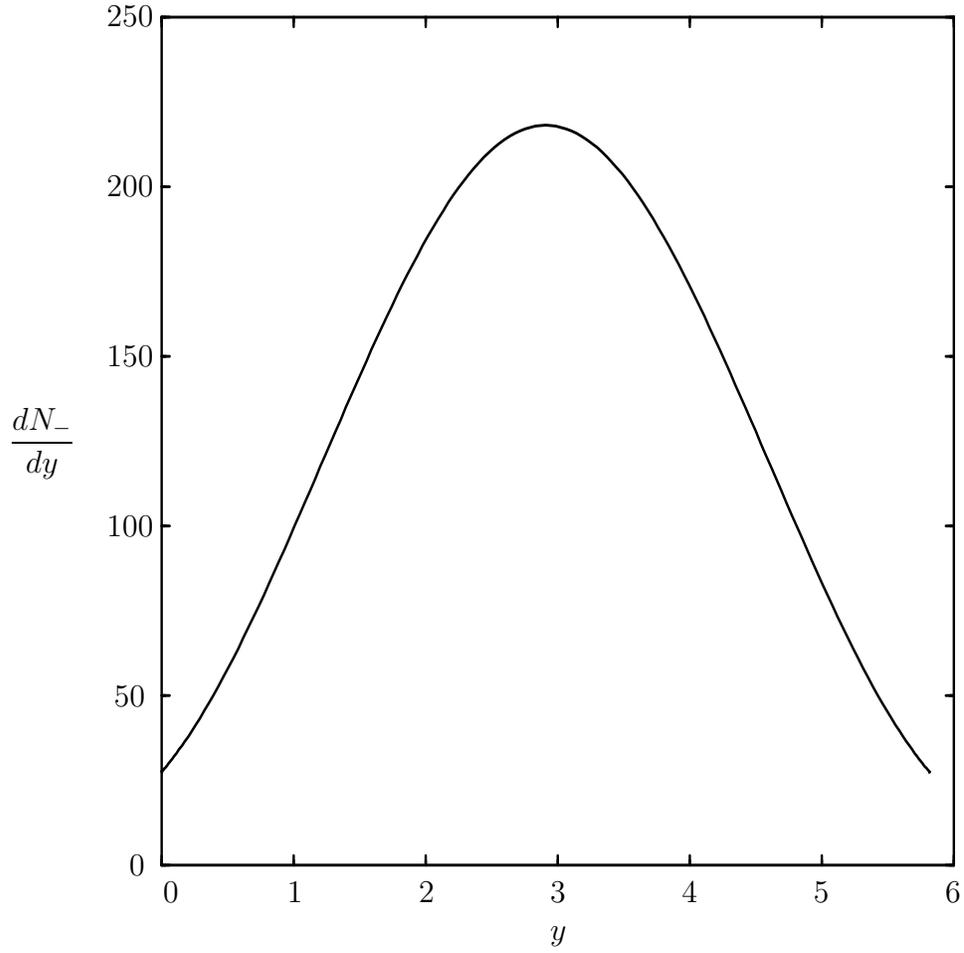,width=12cm,height=12cm}}
\vspace{0.1cm}
\caption{%
LEXUS prediction for the rapidity distribution of negatively charged
hadrons
in 158 GeV$/c$ Pb+Pb collisions with a centrality of 5\%.}
\label{fig:hmdndy_NA49}
\end{figure}

The rapidity distribution of $h^-$ has been measured in S+S collisions
by NA35 \cite{NA35dist}.
The results for the 2\% most central collisions are shown
in Figure~\ref{fig:hmdndy_NA35}.
The prediction of LEXUS is also shown and presents a
very good representation of the data.  Predictions for the 5\% most
central collisions of Pb+Pb at 158 GeV$/c$ per nucleon are shown in
Figure~\ref{fig:hmdndy_NA49}.  No data have yet been published.

\section{Negative Hadron Transverse Momentum Distribution}

Mesons are produced when nucleons collide, and since the colliding
nucleons are undergoing a random walk in transverse momentum, the
mesons will acquire extra transverse momentum too.  This is taken
into account in the following way.

Suppose that the center-of-mass frame takes discrete steps
${\bf v}$ in transverse velocity when baryons
undergo hard scattering.  The magnitude is fixed but its
direction in the transverse plane is random.  Then one can
show that the average transverse momentum
squared of pions produced in a collision between two nucleons which
have together undergone $i$ previous scatterings is
\begin{equation}
\langle p_T^2 \rangle_i = \langle p_T^2 \rangle_{\pi} + \frac{v^2}{1-v^2}
B_i(v^2) \left[ \frac{3}{2} \langle p_T^2 \rangle_{\pi} + m_{\pi}^2 \right]
\end{equation}
where
\begin{equation}
B_i(v^2) = \sum_{k=0}^{i-1} \left( \frac{1+\frac{1}{2}v^2}{1-v^2}
\right)^k \, .
\end{equation}
Here $\langle p_T^2 \rangle_{\pi}$ is the average squared transverse
momentum of pions in elementary nucleon-nucleon collisions.  Identifying
\begin{equation}
\langle p_T^2 \rangle_{NN} = \frac{4m_N^2 v^2}{1-v^2}
\end{equation}
we obtain $v = 0.272$.

The average value of the pion transverse momentum in elementary
nucleon-nucleon collisions varies surprisingly little for
beam momenta ranging from 11.6 GeV$/c$ to 195 GeV$/c$ \cite{Eisen}.
We shall use $\langle p_T^2 \rangle_{\pi} = 0.155\ (\hbox{GeV}/c)^2$.
For pions, too, a good representation of the transverse momentum
distribution in nucleon-nucleon collisions is the thermal form
(\ref{therm}) with the nucleon mass replaced by the pion mass.
This results in a pion temperature in nucleon-nucleon collisions
of 133 MeV.

After $i$ previous scatterings of nucleons the pion distribution
becomes broader with a temperature $T_i$ determined by
\begin{equation}
{\cal N}_i \int_0^{\infty} dp_T\,p_T^3\, m_T\, K_1(m_T/T_i)
=  \langle p_T^2\rangle_i \, .
\end{equation}
The full distribution of pions in rapidity and transverse momentum
is then computed analogously to the procedure in section 5.
\begin{eqnarray}
\frac{d^2N_{\pi^-}}{p_Tdp_T dy} &=&
\sum_{\bar{m}=1}^{A_P} \sum_{\bar{n}=1}^{A_T}
{\cal P}_{\bar n}^T({\bf s}_T) {\cal P}_{\bar m}^P({\bf s}_P)
\sum_{m=1}^{\bar{m}} \sum_{n=1}^{\bar{n}}
\lambda \int dy_P dy_T
W_{mn-1}^P(y_P) W_{m-1n}^T(y_T) \nonumber \\
& \times & \frac{\langle \pi^-(y_{\rm rel}) \rangle_{NN}
\exp\left[-\left(y-y_{\rm cm}\right)^2/2\sigma_L^2(y_{\rm rel})\right]}
{\sqrt{2\pi} \sigma_L(y_{\rm rel})} \nonumber \\
& \times & {\cal N}_{n+m-2} \, m_T\, K_1(m_T/T_{n+m-2}) \, .
\end{eqnarray}
A similar analysis can be done for kaons.  In this case we use
\cite{Eisen} $\langle p_T^2 \rangle_K = 0.290\ (\hbox{GeV}/c)^2$.
We assume that the sum of the $K^-$ and $\pi^-$ multiplicities
equals the $h^-$ multiplicity.

\begin{figure}[p]
\begin{picture}(0,0)(0.1,12)
\put(3.05,0){$0$}
\put(4.7,0){$0.25$}
\put(6.45,0){$0.5$}
\put(8.2,0){$0.75$}
\put(9.95,0){$1.0$}
\put(11.7,0){$1.25$}
\put(13.45,0){$1.5$}
\put(7.7,-0.5){$p_T\ (\hbox{GeV}/c)$}
\put(2.3,0.35){$10^{-2}$}
\put(2.3,2.61){$10^{-1}$}
\put(2.5,4.87){$1$}
\put(2.5,7.13){$10$}
\put(2.3,9.39){$10^2$}
\put(2.3,11.65){$10^3$}
\put(0.8,4.5)
{\begin{sideways}
$\displaystyle {1\over p_T}{dN_{-}\over dp_T}\
\left((\hbox{GeV}/c)^{-2}\right)$
\end{sideways}}
\end{picture}
\centerline{\psfig{figure=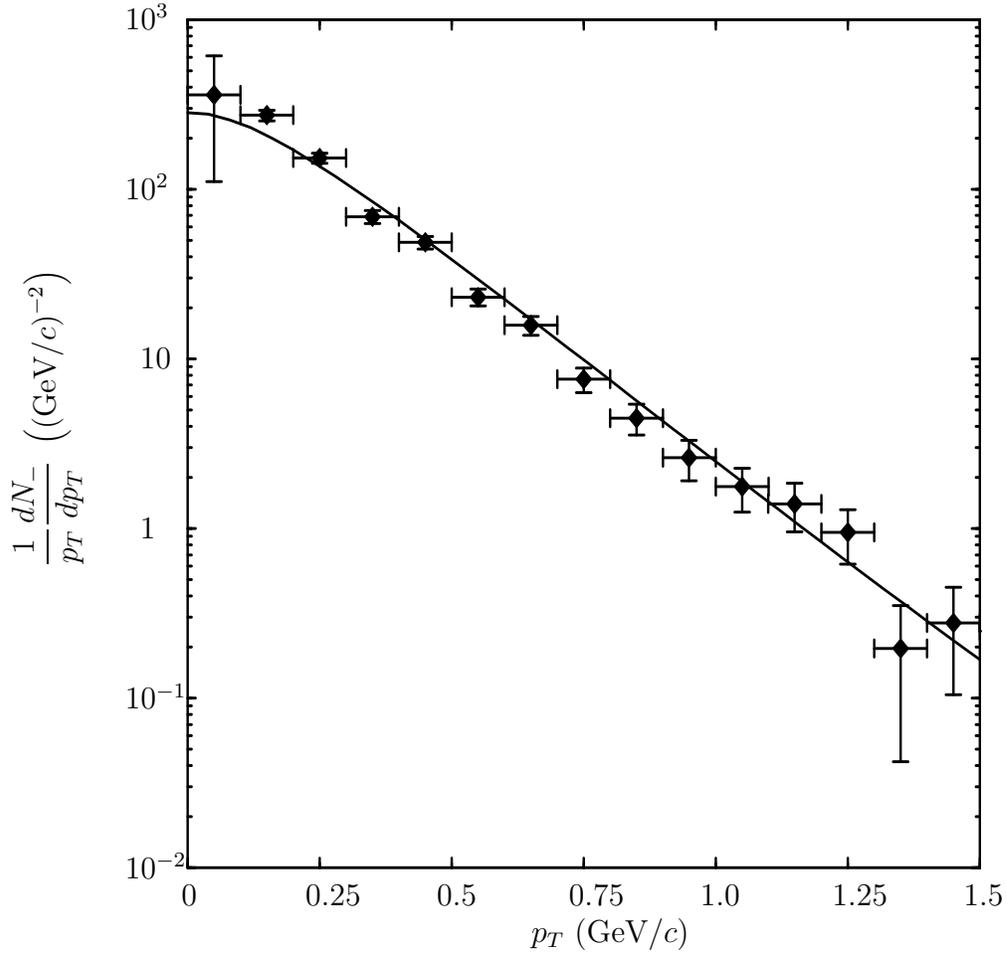,width=12cm,height=12cm}}
\vspace{0.1cm}
\caption{%
Negative hadron transverse momentum distribution for a S+S
collision at 200 GeV$/c$ with a centrality of 2\%.
The solid line represents our result and the filled diamonds are NA35 data.
Rapidity range is $0.8 < y < 2.0$.}
\label{fig:hmpt_08_20_NA35}
\end{figure}

\begin{figure}[p]
\begin{picture}(0,0)(0.1,12)
\put(3.05,0){$0$}
\put(4.7,0){$0.25$}
\put(6.45,0){$0.5$}
\put(8.2,0){$0.75$}
\put(9.95,0){$1.0$}
\put(11.7,0){$1.25$}
\put(13.45,0){$1.5$}
\put(7.7,-0.5){$p_T\ (\hbox{GeV}/c)$}
\put(2.3,0.35){$10^{-2}$}
\put(2.3,2.61){$10^{-1}$}
\put(2.3,4.87){$1$}
\put(2.3,7.13){$10$}
\put(2.3,9.39){$10^2$}
\put(2.3,11.65){$10^3$}
\put(0.8,4.5)
{\begin{sideways}
$\displaystyle {1\over p_T}{dN_{-}\over dp_T}\
\left((\hbox{GeV}/c)^{-2}\right)$
\end{sideways}}
\end{picture}
\centerline{\psfig{figure=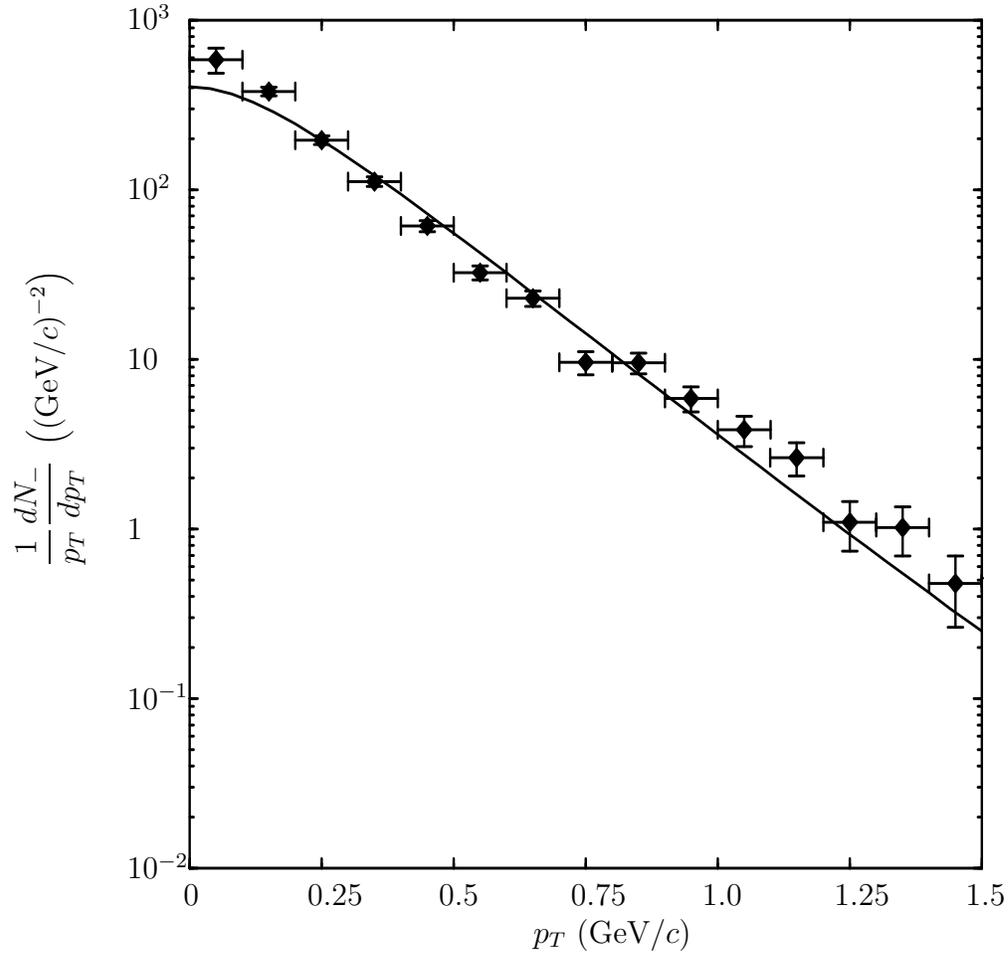,width=12cm,height=12cm}}
\vspace{0.1cm}
\caption{%
The same as figure \protect\ref{fig:hmpt_08_20_NA35} but with
rapidity range $2.0 < y < 3.0$.}
\label{fig:hmpt_20_30_NA35}
\end{figure}

In Figure~\ref{fig:hmpt_08_20_NA35}
we show the $h^-$ transverse momentum distribution
for the 2\% most central S+S collisions for the rapidity interval
$0.8 < y < 2.0$.  The data is from NA35 \cite{NA35dist}.
In Figure~\ref{fig:hmpt_20_30_NA35} we show
the distribution for the interval $2 < y < 3$.  The agreement is
acceptable with no surprises.  The slight underestimate at very small
$p_T$ could be a result of too crude an approximation to the
transverse momentum distribution in nucleon-nucleon collisions.
It could also be a result of multiple scattering among the
produced pions \cite{smallpt}.  We do not attempt to compute the
distribution for $p_T > 1.5$ GeV$/c$ since the parametrization
chosen is not representative of high $p_T$ pion data in
nucleon-nucleon collisions.

\section{Zero Degree Energy Distribution}

Many experiments have what is known as a Zero Degree Calorimeter (ZDC)
which measures the energy carried by forward going particles in a
collision.  It is in some sense a measure of impact parameter since
any projectile nucleons which participate in the collision get
scattered away from the forward direction; to first approximation only
spectator nucleons go into the ZDC.  There is a monotonic relationship
between the average number of spectator nucleons and the impact parameter.
The ZDC is oftentimes used as a centrality trigger whereby acceptance
of only small energy deposition roughly corresponds to central collisions.

The physics of the ZDC is much more complicated than the basic idea
presented above.  The ZDC has a very specific response to a given
hadron ($p$, $n$, $\pi^0$, $K^-$, etc.) with a given LAB momentum.
This response must be accounted for to get an accurate
comparison between a model calculation and the data.  In this paper
we shall make a zero order estimate of the ZDC energy distribution
for heavy ion collisions, and then make a first order correction.

In a heavy ion collision a certain number of projectile nucleons
will not scatter but will continue along the beam direction without
deflection.  Knowing this number we can compute the energy deposited
in the ZDC (see caveat above).  At a given impact parameter the
average number of projectile spectator nucleons is
\begin{equation}
N_{\rm spec}^P(b) =
\int \frac{d^2s_P}{\sigma_{NN}} \sum_{\bar{m}n} {\cal P}^P_{\bar{m}}
({\bf s}_P) \bar{m} {\cal P}^T_n ({\bf s}_T) (1-\lambda)^n \, .
\end{equation}
Here $\bar{m}$ is the number of projectile nucleons in the row
with probability distribution ${\cal P}^P_{\bar{m}}$ and
${\cal P}^T_n (1-\lambda)^n$ is the probability of encountering
$n$ target nucleons without suffering any hard scattering.  All
possibilities are summed over as are all rows.  With the probability
distribution (\ref{binomial}) we get
\begin{eqnarray}
N_{\rm spec}^P(b) &=&
\int \frac{d^2s_P}{\sigma_{NN}} \nu_P({\bf s}_P)
\left[1-\lambda \nu_T({\bf s}_T)/N_{\rm max}\right]^{N_{\rm max}}
\nonumber \\ &\approx&
\int \frac{d^2s_P}{\sigma_{NN}} \nu_P({\bf s}_P)
\exp\left[-\lambda \nu_T({\bf s}_T)\right] \, .
\end{eqnarray}
This has the interpretation of the average number of projectile
nucleons in a row times the probability of not making a hard collision
with any of the target nucleons, integrated over all rows.
When $\lambda \rightarrow 0$ or when $b \rightarrow \infty$
the average number of spectator projectile nucleons approaches
$A_P$ as it should.

There will be fluctuations in the number of spectator nucleons even
at fixed impact parameter.  The dispersion can be computed in the
same way as the average number.
\begin{eqnarray}
D^2_{\rm spec}(b) &=&
\int \frac{d^2s_P}{\sigma_{NN}} \sum_{\bar{m}n} {\cal P}^P_{\bar{m}}
({\bf s}_P) \bar{m} {\cal P}^T_n ({\bf s}_T) (1-\lambda)^n
\left[1-(1-\lambda)^n\right] \nonumber \\
&=& \int \frac{d^2s_P}{\sigma_{NN}} \nu_P({\bf s}_P)
\left\{ \left[1-\lambda \nu_T({\bf s}_T)/N_{\rm max}\right]^{N_{\rm max}}
-\left[1-\lambda (2-\lambda)
\nu_T({\bf s}_T)/N_{\rm max}\right]^{N_{\rm max}} \right\}
\nonumber \\ &\approx&
\int \frac{d^2s_P}{\sigma_{NN}} \nu_P({\bf s}_P)
\left\{ \exp\left[-\lambda \nu_T({\bf s}_T)\right]
- \exp\left[-\lambda (2-\lambda) \nu_T({\bf s}_T)\right] \right\}
\end{eqnarray}
When $\lambda \rightarrow 0$ or when $b \rightarrow \infty$
the dispersion goes to zero, consistent with every single one of
the projectile nucleons entering the ZDC.

We will use the central limit theorem to approximate the conditional
probability distribution of the energy carried by projectile spectators
into the ZDC at fixed impact parameter by a Gaussian.
\begin{equation}
\frac{d{\cal P}}{dE_{\rm ZDC}}(E_{\rm ZDC}|b) = \frac{1}
{\sqrt{2\pi} D(b) E_{\rm beam}}
\exp\left\{ -\left[E_{\rm ZDC}-\langle E(b)\rangle \right]^2
/2D^2(b) E_{\rm beam}^2 \right\}
\label{zdcdis}
\end{equation}
Here we have made the identification $\langle E(b)\rangle
= E_{\rm beam} N_{\rm spec}^P(b)$ and similarly for the dispersion.
To get the ZDC energy distribution we integrate over impact parameter.
\begin{equation}
\frac{d\sigma}{dE_{\rm ZDC}} =
\int d^2b \frac{d{\cal P}}{dE_{\rm ZDC}}(E_{\rm ZDC}|b)
\end{equation}

\begin{figure}[p]
\begin{picture}(0,0)(0,12)
\put(2.95,0){0}
\put(4.25,0){1}
\put(5.5643,0){2}
\put(6.8786,0){3}
\put(8.193,0){4}
\put(9.507,0){5}
\put(10.821,0){6}
\put(12.1357,0){7}
\put(13.45,0){8}
\put(7.5,-0.5){$E_{\rm ZDC}\ (\hbox{TeV})$}
\put(2.1,0.35){$10^{-4}$}
\put(2.1,1.606){$10^{-3}$}
\put(2.1,2.861){$10^{-2}$}
\put(2.1,4.117){$10^{-1}$}
\put(2.1,5.372){$1$}
\put(2.1,6.628){$10$}
\put(2.1,7.883){$10^2$}
\put(2.1,9.139){$10^3$}
\put(2.1,10.39){$10^4$}
\put(2.1,11.65){$10^5$}
\put(0.8,4.5)
{\begin{sideways}
$\displaystyle {d\sigma\over dE_{\rm ZDC}}$ (mb/TeV)
\end{sideways}}
\end{picture}
\centerline{\psfig{figure=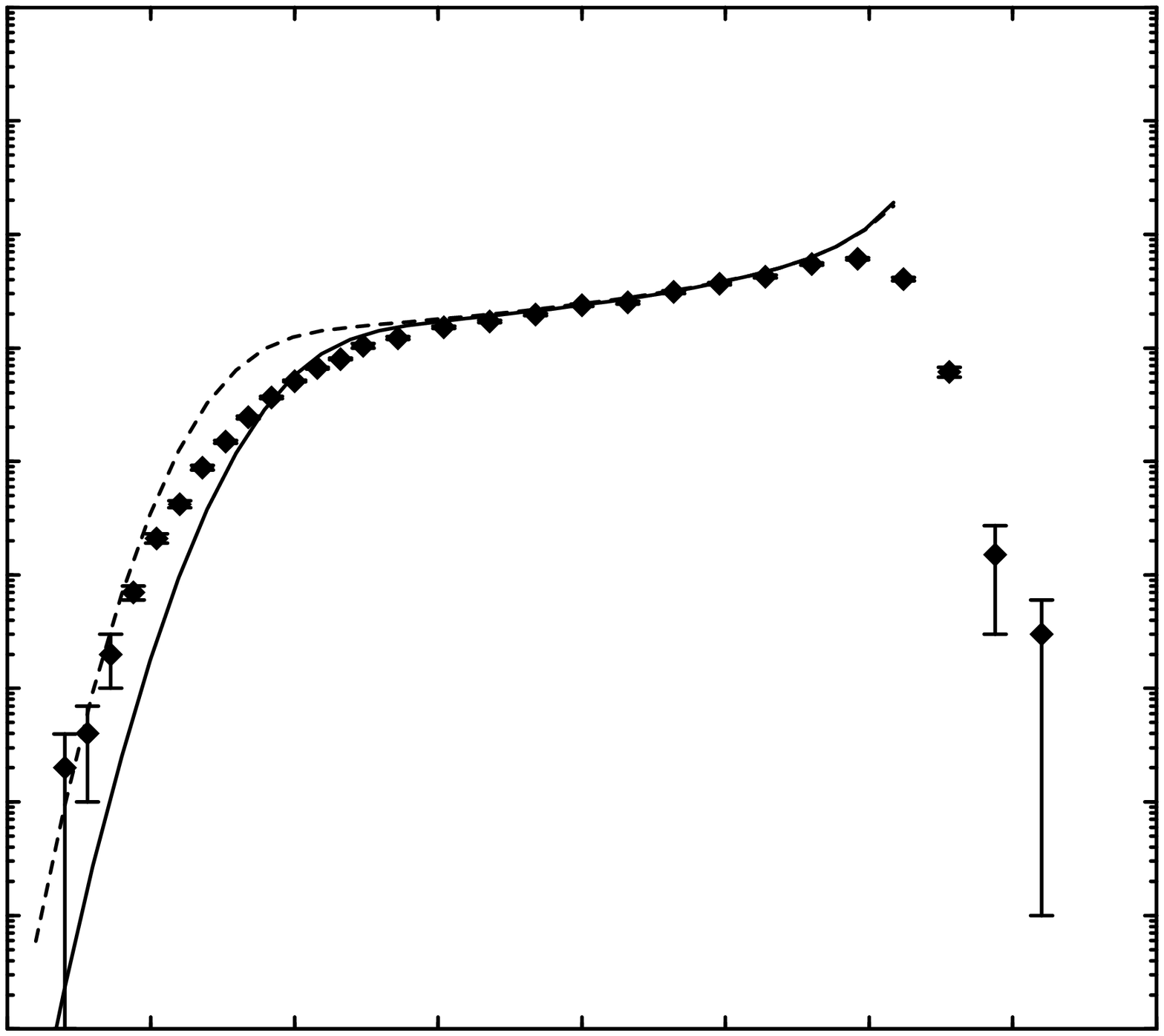,width=12cm,height=12cm}}
\vspace{0.1cm}
\caption{%
The zero degree energy distribution for
200 GeV$/c$ S+S collisions.
The solid line represents LEXUS with an opening angle of
$0.15$ degree.  The dashed line represents LEXUS with only the
spectator nucleons. 
Data are from NA35 \protect\cite{NA35cross}.}
\label{fig:dsdv_NA35}
\end{figure}

Now we compare with experiment.  Data from NA35 for S+S \cite{NA35cross}
is shown in Figure~\ref{fig:dsdv_NA35}
as well as the the cross section for forwardgoing
nucleons in LEXUS (dashed curve).  The central plateau is just about right,
as it should be, reflecting the basic geometry of the nuclei.  The data go
beyond the kinematical limit of 6.4 TeV in a single S+S collision
as a result of inefficiencies in the detector.
LEXUS predicts too much cross section for forwardgoing energies
of 1 to 2 TeV.

\begin{figure}[p]
\begin{picture}(0,0)(0,12)
\put(2.95,0){0}
\put(4.45,0){5}
\put(5.85,0){10}
\put(7.35,0){15}
\put(8.85,0){20}
\put(10.35,0){25}
\put(11.85,0){30}
\put(13.45,0){35}
\put(8.2,-0.5){$E_{\rm ZDC}\ (\hbox{TeV})$}
\put(2.1,0.35){$10^{-2}$}
\put(2.1,2.23333){$10^{-1}$}
\put(2.1,4.11667){$1$}
\put(2.1,6.0){$10$}
\put(2.1,7.88333){$10^{2}$}
\put(2.1,9.76667){$10^{3}$}
\put(2.1,11.65){$10^{4}$}
\put(0.8,4.5)
{\begin{sideways}
$\displaystyle {d\sigma\over dE_{\rm ZDC}}$ (mb/TeV)
\end{sideways}}
\end{picture}
\centerline{\psfig{figure=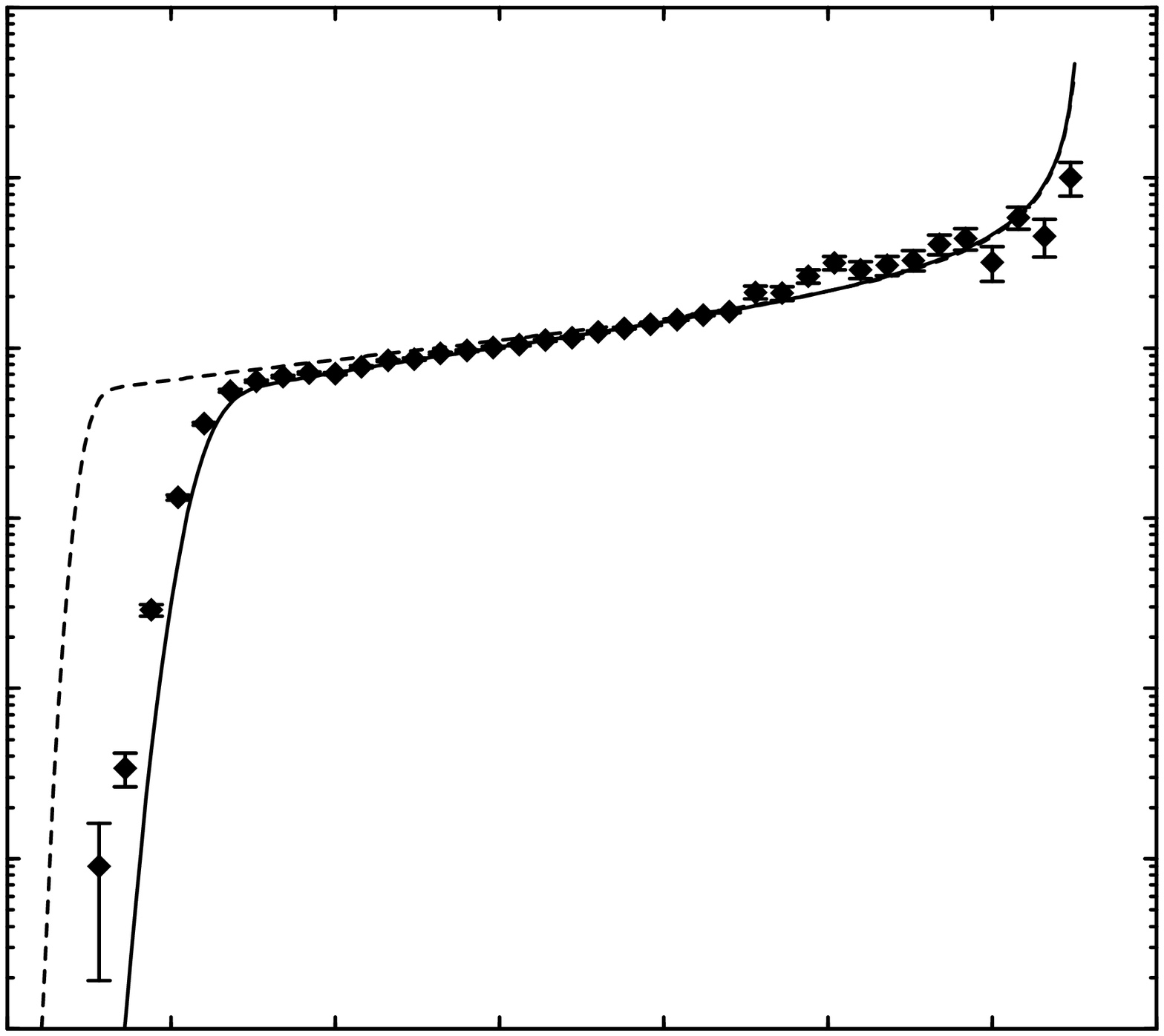,width=12cm,height=12cm}}
\vspace{0.1cm}
\caption{%
The zero degree energy distribution for
158 GeV$/c$ Pb+Pb collisions.
The solid line represents LEXUS with an opening angle of
$0.3$ degree.  The dashed line represents LEXUS with only the
spectator nucleons.
Data are from NA49 \protect\cite{NA49}.}
\label{fig:dsdv_NA49}
\end{figure}

Data from NA49 for Pb+Pb collisions \cite{NA49} is shown
in Figure~\ref{fig:dsdv_NA49}.
Once again the central plateau is correctly reproduced, but the shoulder at
low energy, indicating the most central collisions, is shifted too
far left by about 4 TeV.  This ZDC is improved over that of NA35 and
does not go beyond the kinematic limit of 32.9 TeV.

In actual experiments there is a finite opening angle $\theta_0$
for particle acceptance in the ZDC.  This is generally a fraction
of a degree, in the LAB frame of course.  Some of the hard scattered
nucleons, both projectile and target, may emerge with a LAB angle
smaller than this.  This effect will tend to increase the energy
flow into the ZDC at a fixed impact parameter.  The additional
energy entering the ZDC is
\begin{eqnarray}
E_{0 < \theta < \theta_0}(b) &=& \int dy dp_T \left(m_T \cosh y - m_N\right)
\frac{d^2N^P}{dp_Tdy}(b,p_T,y) \nonumber \\
&\times& \Theta\left(\theta_0 - \tan^{-1}(p_T/m_T \sinh y) \right) \,
\end{eqnarray}
Here $d^2N^P/dp_Tdy$ is the same as expression (\ref{protpt}) but
without the averaging over impact parameter.  Then in eq. (\ref{zdcdis})
we identify $\langle E(b) \rangle = E_{\rm beam} N_{\rm spec}^P(b)
+ E_{0 < \theta < \theta_0}(b)$.  In addition, the dispersion in the
energy entering the ZDC increases and can be estimated by
\begin{equation}
D^2(b)
=
\int {d^2s_P\over \sigma_{NN}}\,
\sum_{\bar{m}n}
{\cal P}_{\bar m}({\bf s}_p)\,
{\cal P}_n({\bf s}_T)\,
\sum_{m=1}^{\bar m}
\sigma^2_{mn}
\;.
\end{equation}
Here
\begin{eqnarray}
\sigma_{mn}^2
&=&
\int dy \int_0^{p_T^{\rm max}}dp_T\,
E_K^2\,
\frac{d^2N^P_{mn}}{dp_Tdy}(p_T,y)\,
-
\left(
\int dy \int_0^{p_T^{\rm max}}dp_T\,
E_K\,
\frac{d^2N^P_{mn}}{dp_Tdy}(p_T,y)
\right)^2
\nonumber\\
& & {}
+ \hbox{Target Contribution}
\end{eqnarray}
where
$E_K = \left(m_T\cosh y - m_N\right)$,
$p_T^{\rm max} = \tan\theta_0 m_T\sinh y$,
and
\begin{eqnarray}
{d^2N^P_{mn}\over p_T dp_T dy}(p_T, y)
& = &
\frac{\overline{W}_{mn}^P(y)}{1-(1-\lambda)^n}
\sum_{k=1}^n
\left( {n \atop k} \right)\,
\lambda^k (1-\lambda)^{n-k}
{\cal N}_k m_T K_1(m_T/T_k)
\nonumber\\
& & {}
+ (1-\lambda)^n\delta(y-y_0)\delta(p_T)
\;.
\end{eqnarray}
The target contribution is obtained by simply changing $y \to y_0-y$.

Now we recompare with experiments.  The results of LEXUS with an
opening angle of 0.15$^o$ is shown in Figure~\ref{fig:dsdv_NA35}
by the solid curve.
The result of allowing some of the scattered nucleons to enter
the ZDC is obvious; it reduces the cross section at small forward
energies.  The results for Pb+Pb with an opening angle of
0.3$^o$ are shown in Figure~\ref{fig:dsdv_NA49}
also with a solid curve.  The better agreement for Pb than for S is
somewhat mysterious.  The opening in the ZDC for both experiments
was actually a square, not a circle, with an aperture of
86 $\mu$sr.  It is quite possible that the simple angle cutoff
we have used does not do justice to the complicated workings of these
calorimeters.  It is also quite possible that our use of a baryon
distribution which factorizes in rapidity and transverse momentum
in nucleon-nucleon collisions is too crude near the edges of phase
space, such as for very forwardgoing nucleons.

\section{Conclusion}

In this paper we have constructed a means to make a linear extrapolation
of nucleon-nucleon collisions to very high energy nucleus-nucleus
collisions.  We call this extrapolation procedure LEXUS.  There is
no reference to quarks, gluons, strings, Pomerons, or QCD.  The
treatment is simply based on a sequence of binary nucleon-nucleon
collisions as in free space.  We know that this treatment cannot
be exact as all of our accumulated knowledge of QCD and high energy
physics of the last twenty-five years can attest.  But it is important
to do these calculations as a baseline against more detailed, but
of necessity approximate, treatments based on perturbative and
nonperturbative QCD to discover thermodynamic properties
of quark-gluon plasma and hadronic matter.  What have we learned?

The rapidity and transverse momentum distributions of baryons in
central sulfur-sulfur and lead-lead collisions at the SPS seem to
be well-described by LEXUS.  The same can be said for the multiplicity
of negatively charged hadrons and their rapidity and transverse momentum
distributions.  The zero degree energy distribution for lead-lead
comes out just about right; for sulfur-sulfur the agreement is less
good.  This may be due to our rather simple treatment of elastic
and diffractive nucleon-nucleon collisions which are more likely to
influence the outcome of the collisions between smaller nuclei.

What is not reproduced so well by LEXUS is the abundance of strange
particles.  Formation of quark-gluon plasma increases the number of
strange quarks in the system due to its relatively small mass in
comparison to the kaon mass \cite{Rafel.1}.  But it is also possible
to increase the number of kaons over that produced in nucleon-nucleon
collisions by multiple scattering and attendant pair production
of kaons \cite{Mek}.  Most notably the number of antilambdas is
too small in LEXUS in comparison to central sulfur-sulfur data
by almost an order of magnitude, indicating that there are proportionately
many more antistrange quarks in the heavy ion collision than in
nucleon collisions \cite{Rafel.2}.  This is interesting physics.
It has been discussed several times at Quark Matter conferences
\cite{qm}.  Our calculations confirm it.  We anxiously await
published data on strangeness in central lead-lead collisions.

There is a preliminary conference report on a forward - backward
azimuthal asymmetry in Pb-Pb collisions \cite{afa}.  This
asymmetry is similar to that seen at much lower Bevalac energies.
It has been interpreted variously as a collective bounce-off,
as if the two nuclei were behaving as fluids, and as the absorption
of particles in the cold spectator matter.  In either way of thinking
the present linear extrapolation of nucleon-nucleon scattering
does not take this crosstalk of rows into consideration.  It would
be challenging to do so \cite{flow}.

A linear extrapolation like LEXUS is only as good as the data
input from nucleon-nucleon collisions.  In this paper we have
used reasonably accurate yet simple parametrizations of the
basic input data.  Nevertheless improvements can be made.  For
example, our treatment of the elastic and diffractive components
of nucleon-nucleon collisions could be represented more accurately
but at the cost of significant complication to the solution to the
model as described in section 2.  A better prediction for the
baryon momentum distribution near the projectile and target rapidities
would likely result, as would a description of the energy deposited
in a zero degree calorimeter.  Even then, our ability to make
a linear extrapolation will be hindered by the lack of experimental
measurements of many-particle correlations, such as between the
two outgoing baryons in an elementary collision, or the correlation
between the rapidities of outgoing baryons with the number of
produced mesons.  It is unlikely that all the exclusive cross
sections for nucleon-nucleon, nucleon-hyperon, and hyperon-hyperon
collisions at all the energies of relevance will ever be known
experimentally.  This is a particular shortcoming if one wants
to make a Monte Carlo event generator out of LEXUS.  In case
of improvements to LEXUS in the future the model described in
this paper will be known as version 1.0.

We have not made any comparison to transverse energy distributions
as measured by electromagnetic calorimeters.  Many such measurements
have been made, but it is a challenging task to normalize them in
such a way that we can know how much energy each type of hadron
($p$, $n$, $\pi^0$, $K^-$, etc.) actually deposits to within a
certain accuracy to make a comparison meaningful.  This will be
the subject of a future application of LEXUS.

Also under study is an application of LEXUS to some very interesting
results on the production of $J/\psi$, photons, and lepton pairs
as discussed in recent Quark Matter conferences \cite{qm}.

Finally, it would be interesting to make LEXUS predictions for
RHIC where we certainly anticipate it to fail, lacking a
description of hard perturbative QCD \cite{Geiger,Wang} and
the inside-outside cascade effect \cite{BJ}.  The difficulty
is that there is no nucleon-nucleon data available at RHIC energy
before the turn-on of the accelerator.  Nevertheless a prediction
for the baryon rapidity distribution may be considered and is
now in progress.

\section*{Acknowledgements}

We thank
Marek Ga\'zdzicki,
John Harris,
Peter Jacobs,
Spiros Margetis,
Milton Toy,
Flemming Videbaek
and
Nu Xu
for helpful discussions of the experiments and data.
We also thank Laszlo Csernai for helpful comments and discussions. 
This work was supported by the US Department of Energy
under grant DE-FG02-87ER40328.

\newpage

\begin{table}[h]
\begin{center}
\begin{tabular}{|l||c|c|c|c|c|}
\hline
\vphantom{$\Big($}
    & $K^+$  & $K^-$   & $K^0_S$ & $\Lambda$ & $\bar{\Lambda}$
\\
\hline
\vphantom{$\Big($}
$a$ & 0.208 & 0.0190 & 0.0426 & 0.0891 & 0.00203
\\
\hline
\vphantom{$\Big($}
$b$ & 6.92 & 2.62   & 4.28   & 4.03   & 5.67
\\
\hline
\vphantom{$\Big($}
$c$ & 2.24  & 2.20  & 0.794  & 1.16   & 2.14
\\
\hline
\end{tabular}
\end{center}
\caption{
Table of the parameters for the fits in
eqs.~(\ref{eq:fit1}--\ref{eq:fit2}).}
\label{tbl:para1}
\end{table}

\begin{table}[h]
\begin{center}
\begin{tabular}{|l||c|c|c|c|c|c|}
\hline
\vphantom{$\Big($}
    & $h^-$ & $K^+$  & $K^-$   & $K^0_S$ & $\Lambda$ & $\bar{\Lambda}$
\\
\hline
\vphantom{$\Big($}
NA35
& $98\pm 3$
& $12.5\pm 0.4$
& $6.9\pm 0.4$
& $10.5\pm 1.7$
& $9.4\pm 1.0$
& $2.2\pm 0.4$
\\
\hline
\vphantom{$\Big($}
LEXUS $0\%$
& 106
& 9.7 $^{+1.5}_{-2.4}$
& 5.7 $^{+0.8}_{-1.4}$
& 5.2 $^{+0.2}_{-0.8}$
& 4.1 $^{+0.5}_{-2.1}$
& 0.29 $^{+0.46}_{-0.17}$
\\
\hline
\vphantom{$\Big($}
LEXUS $2\%$
& 102
& 9.4 $^{+1.4}_{-2.3}$
& 5.5 $^{+0.8}_{-1.3}$
& 5.0 $^{+0.2}_{-0.8}$
& 3.9 $^{+0.5}_{-2.0}$
& 0.28 $^{+0.44}_{-0.16}$
\\
\hline
\vphantom{$\Big($}
LEXUS $4\%$
& 97.8
& 9.0 $^{+1.4}_{-2.2}$
& 5.3 $^{+0.8}_{-1.3}$
& 4.8 $^{+0.2}_{-0.7}$
& 3.8 $^{+0.4}_{-2.0}$
& 0.27 $^{+0.43}_{-0.16}$
\\
\hline
\end{tabular}
\end{center}
\caption{
The average particle multiplicities in S+S collisions.  The NA35 data
should be compared against the 2\% most central collisions in LEXUS.}
\label{tbl:mult}
\end{table}

\begin{table}[h]
\begin{center}
\begin{tabular}{|l||c|c|c|c|c|c|}
\hline
\vphantom{$\Big($}
    & $h^-$ & $K^+$  & $K^-$   & $K^0_S$ & $\Lambda$ & $\bar{\Lambda}$
\\
\hline
LEXUS $0\%$
\vphantom{$\Big($}
& 886
& 68 $^{+13}_{-16}$
& 37 $^{+6}_{-10}$
& 35 $^{+0.5}_{-3}$
& 33 $^{+3}_{-17}$
& 1.6 $^{+2.0}_{-0.9}$
\\
\hline
LEXUS $5\%$
\vphantom{$\Big($}
& 781
& 60 $^{+11}_{-14}$
& 32 $^{+6}_{-9}$
& 31 $^{+0.5}_{-3}$
& 29 $^{+3}_{-15}$
& 1.4 $^{+1.8}_{-0.8}$
\\
\hline
\end{tabular}
\end{center}
\caption{
Predictions for 158 GeV$/c$ Pb+Pb collisions assuming 0 and 5\% centrality.
}
\label{tbl:mult_Pb}
\end{table}

\clearpage

\clearpage

\clearpage

\clearpage

\clearpage

\clearpage

\clearpage

\clearpage

\clearpage

\clearpage

\clearpage

\clearpage

\clearpage

\clearpage

\clearpage


\begin{thebibliography} {99}
\bibitem{mass} This reduces to the distribution $Q$ used in \cite{Hwa}
and \cite{Csernai} if the nucleon mass is neglected at high momentum.
\bibitem{web} To be made available after publication of this paper.
\bibitem{temp} By temperature we mean a parameter characterizing a
distribution.  We make no connection to thermodynamics in this paper.
\end{thebibliography}
\end{document}